\documentclass[12pt,leqno]{article}

\newcommand{\lrej}{\ensuremath{{L_{rej}}}}
\newcommand{\boldturingred}{\mbox{\boldmath${{\,\leq_{\rm T}^{{\rm p}}\,}}$}}
\newcommand{\snred}{\ensuremath{\,\leq_{\rm T}^{{\rm SN}}\,}}
\newcommand{\snredNoCommaspace}{\ensuremath{\leq_{\rm T}^{{\rm SN}}}}
\newcommand{\boldsnred}{\mbox{\boldmath${{\,\leq_{\bf T}^{{\bf SN}}\,}}$}}
\newcommand{\rsred}{\ensuremath{\,\leq_{\rm T}^{{\rm RS}}\,}}
\newcommand{\notrsred}{\ensuremath{\,\not\leq_{\rm T}^{{\rm RS}}\,}}
\newcommand{\rsredNoCommaspace}{\ensuremath{\leq_{\rm T}^{{\rm RS}}}}
\newcommand{\boldrsred}{\mbox{\boldmath${{\,\leq_{\bf T}^{{\bf RS}}\,}}$}}
\newcommand{\ored}{\ensuremath{\,\leq_{\rm T}^{{\rm O}}\,}}
\newcommand{\boldored}{\mbox{\boldmath${{\,\leq_{\bf T}^{{\bf O}}\,}}$}}
\newcommand{\ured}{\ensuremath{\,\leq_{\rm T}^{{\rm U}}\,}}
\newcommand{\boldured}{\mbox{\boldmath${{\,\leq_{\bf T}^{{\bf U}}\,}}$}}
\newcommand{\npinterconp}{{\rm NP} \cap {\rm coNP}}
\newcommand{\dann}{\longrightarrow}
\newcommand{\tm}{\subseteq}
\newcommand{\et}{{\rm EXP}}
\newcommand{\gdw}{\longleftrightarrow}

\usepackage{xspace}

\usepackage{latexsym}

\let\ensuremathTEMP\ensuremath

\makeatletter%
\def\nottoobig#1{{\hbox{$\left#1\vcenter to1.111\ht\strutbox{}\right.\n@space$}}}
\makeatother%

\makeatother%

\makeatletter

\newcount\hour  \newcount\minutes  \hour=\time  \divide\hour by 60
\minutes=\hour  \multiply\minutes by -60  \advance\minutes by \time
\def\mmmddyyyy{\ifcase\month\or Jan\or Feb\or Mar\or Apr\or May\or Jun\or Jul\or
  Aug\or Sep\or Oct\or Nov\or Dec\fi \space\number\day, \number\year}
\def\hhmm{\ifnum\hour<10 0\fi\number\hour :%
  \ifnum\minutes<10 0\fi\number\minutes}
\def\Draft{{\it Draft of \mmmddyyyy}}

\def\ps@jtsheadings{%
\def\@oddhead{\it\rightmark\hfil\rm\thepage}%
\def\@oddfoot{\hfil\Draft}%
\if@twoside
\def\@evenhead{\rm\thepage\hfil\it\leftmark}%
\def\@evenfoot{\Draft\hfil}%
\else
\let\@evenhead\@oddhead%
\let\@evenfoot\@oddfoot%
\fi
}
\def\ps@jtsplain{%
\def\@oddhead{\hfil\Draft}%
\def\@oddfoot{\hfil\rm\thepage\hfil}%
\let\@evenfoot\@oddfoot%
\if@twoside \def\@evenhead{\Draft\hfil} \else \let\@evenhead\@oddhead \fi
}

\def\chaptermark#1{\markboth{\thechapter.\ #1}{\thechapter.\ #1}}%
\def\sectionmark#1{\markright{\thesection.\ #1}}

\def\section{\@startsection {section}{1}{\z@}
    {3.5ex plus1ex minus.2ex}{2.3ex plus.2ex}{\Large\bf}}
\def\subsection{\@startsection{subsection}{2}{\z@}
    {3.25ex plus1ex minus.2ex}{1.5ex plus.2ex}{\large\bf}}
\def\subsubsection{\@startsection{subsubsection}{3}{\z@}
    {3.25ex plus1ex minus.2ex}{1.5ex plus.2ex}{\normalsize\bf}}
\def\paragraph{\@startsection{paragraph}{4}{\z@}
    {3.25ex plus1ex minus.2ex}{1em}{\normalsize\bf}}
\def\subparagraph{\@startsection{subparagraph}{4}{\parindent}
    {3.25ex plus1ex minus.2ex}{1em}{\normalsize\bf}}

\makeatother

\makeatletter \@beginparpenalty=10000 \makeatother
\def\underl#1 {\leavevmode\let\first=\relax\underli #1 }
\def\underli#1 {\ifx&#1\let\next=\relax\unskip
                \else\let\next=\underli\first\ulinebox{#1}\fi\let\first=\undersp\next}
\def\undersp{\penalty50\ulinebox{\space}\penalty50}
\def\ulinebox#1{\vtop{\hbox{\strut#1}\hrule}}
\def\unice#1 {\underl #1 & }

\def\desclabel#1{\bf #1\hfil}
\def\desc{\list{}{%
\labelwidth= \leftmargin
\advance \labelwidth by -\labelsep
\let \makelabel=\desclabel}}



\makeatletter 

\newcommand{\oldimplies}{\:\Rightarrow\:}

\def\inter{\,\cap\,}
\newcommand{\true}{\mbox{\it true}}
\newcommand{\false}{\mbox{\it false}}

\newlength{\filength}
\newsavebox{\gcbox}
\sbox{\gcbox}{\framebox[\filength]{\rule{0ex}{2ex}}}

\newlength{\leftjustindent}
\newlength{\@leftjustindent}
\def\leftjust{\let\\\@leftjustcr\let\end\@endleftjust
  \addtolength{\@leftjustindent}{\leftjustindent} \vcenter\bgroup
\halign\bgroup \hbox to\displaywidth{
\rule{\@leftjustindent}{0ex}$\displaystyle##$\hfill }\crcr }
\def\endleftjust{\crcr\egroup\egroup\endgroup}
\def\@endleftjust#1{\crcr\egroup\egroup\@checkend{#1}\endgroup}
\def\@leftjustcr{\crcr}

\newcommand{\red}[3]{ {  {\rm R}_{#1}^{#2}({#3}) }    }
\newcommand{\sparse}{{{\rm SPARSE}}}

\newtheorem{theorem}{Theorem}[section]
\newtheorem{corollary}[theorem]{Corollary}

\newtheorem{cond}[theorem]{Condition}
\newtheorem{question}[theorem]{Question}

\newcommand{\qedblob}{\mbox{$\Box$}}
\def\literalqed{{\ \nolinebreak\hfill\mbox{\qedblob\quad}}}

\def\qed{\literalqed}

\newtheorem{proposition}[theorem]{Proposition}
\newcommand{\singlespacing}{\let\CS=
\@currsize\renewcommand{\baselinestretch}{1}\tiny\CS}
\newcommand{\singlespacingplus}{\let\CS=
\@currsize\renewcommand{\baselinestretch}{1.25}\tiny\CS}
\newcommand{\doublespacing}{\let\CS=
\@currsize\renewcommand{\baselinestretch}{1.75}\tiny\CS}
\newcommand{\draftspacing}{\let\CS=
\@currsize\renewcommand{\baselinestretch}{2.0}\tiny\CS}
\newcommand{\normalspacing}{\singlespacing}
\makeatother

\mathcode`\0="0030      
\mathcode`\1="0031
\mathcode`\2="0032
\mathcode`\3="0033
\mathcode`\4="0034
\mathcode`\5="0035
\mathcode`\6="0036
\mathcode`\7="0037
\mathcode`\8="0038
\mathcode`\9="0039
\singlespacingplus

\newtheorem{definition}[theorem]{Definition}
\newtheorem{notation}[theorem]{Notation}

\makeatletter
\clubpenalty=\@highpenalty
\widowpenalty=\@highpenalty
\makeatother

\makeatletter
\newcommand{\niceonespacing}{\let\CS=\@currsize\renewcommand{\baselinestretch}{1.1}\tiny\CS}\newcommand{\nicetwospacing}{\let\CS=\@currsize\renewcommand{\baselinestretch}{1.2}\tiny\CS}
\newcommand{\nicethreespacing}{\let\CS=\@currsize\renewcommand{\baselinestretch}{1.3}\tiny\CS}
\newcommand{\nicefoospacing}{\let\CS=\@currsize\renewcommand{\baselinestretch}{1.2}\tiny\CS}
\newcommand{\singlespacingplusplus}{\let\CS=\@currsize\renewcommand{\baselinestretch}{1.35}\tiny\CS}
\newcommand{\nicefourspacing}{\let\CS=\@currsize\renewcommand{\baselinestretch}{1.4}\tiny\CS}
\newcommand{\nicefivespacing}{\let\CS=\@currsize\renewcommand{\baselinestretch}{1.5}\tiny\CS}
\newcommand{\nicesixspacing}{\let\CS=\@currsize\renewcommand{\baselinestretch}{1.6}\tiny\CS}
\makeatother

\makeatletter
\def\@cite#1#2{[#1\if@tempswa , #2\fi]}
\makeatother

\makeatletter
\def\@citex[#1]#2{\if@filesw\immediate\write\@auxout{\string\citation{#2}}\fi
  \def\@citea{}\@cite{\@for\@citeb:=#2\do
    {\@citea\def\@citea{,\linebreak[0]}\@ifundefined
       {b@\@citeb}{{\bf ?}\@warning
       {Citation `\@citeb' on page \thepage \space undefined}}%
\hbox{\csname b@\@citeb\endcsname}}}{#1}}
\makeatother

\makeatletter
\def\ps@thesis{\def\@oddhead{\hfil\rm\thepage\hfil}\def\@oddfoot{}\def\@evenhead{\hfil\rm\thepage\hfil}\def\@evenfoot{}\def\chaptermark##1{}\def\sectionmark##1{}}
\makeatother

\makeatletter
\def\foobarpt{\textfont\z@\tenrm 
  \scriptfont\z@\ninrm \scriptscriptfont\z@\sevrm
\textfont\@ne\tenmi \scriptfont\@ne\ninmi \scriptscriptfont\@ne\sevmi
\textfont\tw@\tensy \scriptfont\tw@\ninsy \scriptscriptfont\tw@\sevsy
\textfont\thr@@\tenex \scriptfont\thr@@\tenex \scriptscriptfont\thr@@\tenex
\def\unboldmath{\everymath{}\everydisplay{}\@nomath\unboldmath
          \textfont\@ne\tenmi 
          \textfont\tw@\tensy \textfont\lyfam\tenly
          \@boldfalse}\@boldfalse
\def\boldmath{\@ifundefined{tenmib}{\global\font\tenmib\@mbi\@magscale1\global
        \font\tensyb\@mbsy \@magscale1\global\font
         \tenlyb\@lasyb\@magscale1\relax\@addfontinfo\@xiipt
              {\def\boldmath{\everymath
                {\mit}\everydisplay{\mit}\@prtct\@nomathbold
                \textfont\@ne\tenmib \textfont\tw@\tensyb 
                \textfont\lyfam\tenlyb\@prtct\@boldtrue}}}{}\@xiipt\boldmath}%
\def\prm{\fam\z@\tenrm}%
\def\pit{\fam\itfam\tenit}\textfont\itfam\tenit \scriptfont\itfam\ninit
   \scriptscriptfont\itfam\sevit
\def\psl{\fam\slfam\tensl}\textfont\slfam\tensl 
     \scriptfont\slfam\tensl \scriptscriptfont\slfam\tensl
\def\pbf{\fam\bffam\tenbf}\textfont\bffam\tenbf 
   \scriptfont\bffam\ninbf \scriptscriptfont\bffam\ninbf 
\def\ptt{\fam\ttfam\tentt}\textfont\ttfam\tentt
   \scriptfont\ttfam\nintt \scriptscriptfont\ttfam\nintt 
\def\psf{\fam\sffam\tensf}\textfont\sffam\tensf
    \scriptfont\sffam\tensf \scriptscriptfont\sffam\tensf
\def\psc{\@getfont\psc\scfam\@xiipt{\@mcsc\@magscale1}}%
\def\ly{\fam\lyfam\tenly}\textfont\lyfam\tenly 
   \scriptfont\lyfam\ninly \scriptscriptfont\lyfam\sevly
 \@setstrut \rm}

\makeatother

\newcommand{\coding}{{\rm coding}}
\newcommand{\outcome}{{\rm outcome}}

\newcommand{\sat}{{\rm SAT}}

\newcommand{\p}{{\rm P}}
\newcommand{\littlep}{{p}}

\newcommand{\np}{{\rm NP}}

\newcommand{\conp}{{\rm coNP}}

\newcommand{\pitwo}{\ensuremath{\Pi_2^{\rm p}}}
\newcommand{\thetatwo}{\ensuremath{\Theta_2^{\rm p}}}

\newcommand{\poly}{\ensuremath{{\rm poly}}}

\newcommand{\ph}{\ensuremath{{\rm PH}}}

\newcommand{\sproof}{\noindent{\bf Proof}\quad}

\newcommand{\dtime}{\ensuremath{{\rm DTIME}}}

\newcommand{\subsetproper}{  \stackrel{\scriptscriptstyle\subset}{\scriptscriptstyle\not-}}

\normalspacing

\newcommand{\pairs}[1]{\mathopen\langle{#1}\mathclose\rangle}

\newcommand{\turingred}{\ensuremath{\,\leq_{\rm T}^{\rm {\littlep}}\,}}

\newcommand{\sigmastar}{\ensuremath{\Sigma^\ast}}
\newcommand{\pisnp}{\ensuremath{\p=\np}}

\newcommand{\pisnotnp}{\ensuremath{\p\neq\np}}

\newcommand{\calc}{\ensuremath{{\cal C}}}

\newcommand{\bigo}{{\protect\cal O}}

\newcommand{\condition}{\,\nottoobig{|}\:}
 
\def\land{{\; \wedge \;}}
\newcommand{\notored}{\not\le^{\rm O}_{\rm T}}
\newcommand{\notured}{\not\le^{\rm U}_{\rm T}}

\normalspacing
\title{Robust Reductions}

\author{
{\em  Jin-Yi Cai\/}\protect\thanks{Research supported in part by
grants NSF-CCR-9057486 and
NSF-CCR-9319093, and 
an Alfred P.~Sloan Fellowship.}  \\
  Department of Computer Science\\
State University of New York at Buffalo\\
             Buffalo, NY 14260, USA
\and
{\em  Lane A. Hemaspaandra\/}\protect\thanks{Supported
in part by grants
NSF-CCR-9322513,
NSF-INT-9513368/DAAD-315-PRO-fo-ab,
and NSF-INT-9815095/DAAD-315-PPP-g\"u-ab.  Work done in part 
while visiting Friedrich-Schiller-Universit\"at.}
\\Department of Computer Science\\University of Rochester\\
             Rochester, NY 14627, USA
\and
{\em  Gerd Wechsung\/}\protect\thanks{Supported
in part by grants
NSF-INT-9513368/DAAD-315-PRO-fo-ab
and 
NSF-INT-9815095/DAAD-315-PPP-g\"u-ab. 
Work done in 
part while visiting Le~Moyne College.}
\\Institut f\protect\"ur Informatik \\
        Friedrich-Schiller-Universit\protect\"at Jena\\
        07740 Jena, Germany
}
\date{}

\makeatletter
\def\@listI{\leftmargin\leftmargini \parsep 4.5pt plus 1pt minus 1pt\topsep
6pt plus 2pt minus 2pt \itemsep  2pt plus 2pt minus 1pt}

\let\@listi\@listI
\@listi
\makeatother

\newcommand{\zpp}{{\rm ZPP}}

\newcommand{\zppnp}{{\zpp^{\rm NP}}}

\makeatletter 
 \newcommand{\setoffdisplay}{\rule{5.9in}{1pt}}

\makeatother



\newcommand{\fp}{\ensuremathTEMP{{\rm FP}}}

\newcommand{\sigmazero}{\ensuremathTEMP{{\Sigma_0^{\rm p}}}}

\newcommand{\sigmatwo}{\ensuremathTEMP{{\Sigma_2^{\rm p}}}}

\newcommand{\sigmaiplusone}{\ensuremathTEMP{{\Sigma_{i+1}^{\rm p}}}}

\newcommand{\pii}{\ensuremathTEMP{{\Pi_{i}^{\rm p}}}}
\newcommand{\sigmai}{\ensuremathTEMP{{\Sigma_{i}^{\rm p}}}}

\begin{document}

\typeout{WARNING:  BADNESS used to suppress reporting.  Beware.}
\hbadness=3000
\vbadness=10000 

\bibliographystyle{plain}

\pagestyle{empty}



{
\singlespacing

\maketitle

}


{\singlespacing

\begin{abstract}
We continue the study of  robust reductions initiated
by Gavald\`{a} and Balc\'{a}zar.
In particular, a 1991 paper of
Gavald\`{a} and Balc\'{a}zar~\cite{bal-gav:j:rob}
claimed
an optimal separation between the power
of robust and nondeterministic strong reductions.
Unfortunately, their proof is invalid.
We re-establish
their theorem.

Generalizing robust reductions, we note that robustly strong reductions are
built from two restrictions, robust underproductivity and robust
overproductivity, both of which have been separately studied before in other
contexts. By systematically analyzing the power of these reductions,
we explore the extent to which each restriction
weakens the power of reductions.  We show that 
one of these reductions yields
a new, strong form of the 
Karp-Lipton Theorem.
\end{abstract}

}  


\normalspacing

\pagestyle{plain}
\sloppy


\section{Introduction}\label{sec:intro}
The study of the relative power of reductions has long been one of
central importance in computational complexity theory. Reductions are
the key tools used in complexity theory to
compare the difficulty of problems.  When we say that $A$
reduces to $B$, we informally interpret this to mean ``$A$ is roughly
easier than $B$,'' where the ``roughly'' regards a certain tolerance that
reflects the power or flexibility of the reduction. To understand
 precisely the  complexity of a problem, we must  understand
the nature of this tolerance, and 
thus we must understand the relative power of
reductions.

Beyond that, reductions play a central role in countless theorems of
complexity theory, and to understand the power of such theorems
we must understand the relationships between reductions.  For example,
Karp and Lipton~\cite{kar-lip:c:nonuniform} proved that if SAT
Turing-reduces to some sparse set then the polynomial hierarchy
collapses.  A more careful analysis
reveals that the same result applies  under the weaker
hypothesis that SAT robustly-strong-reduces to some sparse set.  In
fact, the latter result is simply a relativized version of the former
result~\cite{hem-hoe-nai-ogi-sel-thi-wan:j:np-selective}, though the
first proofs of the latter result were direct and 
quite complex~\cite{aba-fei-kil:j:hide,kae:j:nup}.  
As another example, in the
present paper---but not by simply asserting
relativization---we will note that various theorems,
among them the Karp-Lipton Theorem,
indeed hold for certain reductions
that are even more flexible than robustly strong reductions.

In this paper, we continue the investigation 
of
robust
reductions started 
by
Gavald\`{a} and Balc\'{a}zar~\cite{bal-gav:j:rob}.
We now briefly mention one way of defining 
strong 
reduction~\cite{sel:j:enumeration-reducibility,lon:j:nondeterministic-reducibilities}
and robustly strong 
reduction~\cite{bal-gav:j:rob}.
Definition~\ref{def:twored} provides a formal definition
of the same notions in terms of concepts that are central
to this paper.
We say that a nondeterministic Turing machine is a 
nondeterministic polynomial-time Turing machine 
(NPTM) if 
there is a polynomial $p$ such that, for each oracle $A$
and for each integer $n$, 
the nondeterministic runtime of $N^A$ on inputs of 
size $n$ is bounded by $p(n)$.  (Requiring
that the polynomial upper-bounds the 
runtime {\em independent of the oracle\/} is superfluous in
the definition of $\snredNoCommaspace$, but may be a nontrivial
restriction in the definition of $\rsredNoCommaspace$;
see the 
discussion of this point in 
Section~\ref{s:local-vs-global}.  The 
definitions used here agree with those in
the previous literature.)
Consider 
NPTMs with
three possible outcomes on each path: {\bf acc}, {\bf rej}, 
and~{\bf{}?}.
We say $A$ {\it strong-reduces to} $B$, $A \snred B$,
if there is an NPTM $N$ such that, for every input $x$,
it holds that (a)~if 
$x \in A$ then $N^B(x)$ has at least one {\bf acc} path
and no {\bf rej} paths,
and (b)~if $x \not\in A$ then $N^B(x)$ has at least
one {\bf rej} path
and no {\bf acc} paths. (Note that 
in either case the machine may also have some~{\bf{}?}~paths.)
Furthermore, we say
$A$ {\it robustly strong-reduces to} $B$, $A \rsred B$,
if there is an NPTM $N$ such that $A \snred B$ via $N$
(in the sense of the above definition)
and, moreover, for every oracle $O$ and every input $x$,
$N^O(x)$ is {\it strong}, i.e., it
either has at least one {\bf acc} path and no {\bf rej} paths,
or has at least one {\bf rej} path and no {\bf acc} paths. 
This paper is concerned
with the relative power of these two reductions,
and with reductions whose power is intermediate between theirs.

In particular,  it is claimed in \cite{bal-gav:j:rob} that the following
strong separation holds with respect to the two reductions:

{\singlespacing

\begin{quote}
For every recursive set $A \not\in \np \inter \conp$, there is a
 recursive set $B$ such that $A$ strong-reduces to $B$ but $A$
does not robustly strong-reduce to
$B$~\cite[Theorem~11]{bal-gav:j:rob}.
\end{quote}

}

Unfortunately, there is a subtle but apparently fatal error
in their proof. One of the main contributions of this paper is that
we re-establish their sweeping theorem. Note that the zero degrees
of these reducibilities are identical, namely the class
$\np \inter \conp$~\cite{bal-gav:j:rob}.
Thus, in a certain sense,
the above claim of Gavald\`{a} and Balc\'{a}zar is optimal (if it 
is true, as we prove it is), as if 
$A \in \np \inter \conp$  then 
$A$ strong-reduces to every $B$ and  $A$
also robustly strong-reduces to every
$B$.

Section~\ref{sec:xy} 
presents our proof of the above claim of Gavald\`{a} and Balc\'{a}zar.
The proof is delicate, and is carried out
in three stages:  First, we establish the result
for all $A \in \et - (\np \inter \conp)$, 
where $\et = \cup_{k>0\,}\dtime[2^{n^k}]$. Here, the set $B$
produced from the proof is not necessarily recursive.  Second, we
remove the restriction of $A \in \et$, by showing that
if the result fails for $A \not\in\et$ then indeed $A \in \et$,
yielding a contradiction.  
The proof so far only establishes the existence of 
some $B$, which is not necessarily recursive.  Finally,
with the certainty that {\it some} $B$ exists, we can recast the proof
and show that
for every recursive $A$ a recursive $B$ can be constructed.
We mention to the reader the referee's comment 
that~\cite{gil-sim:c:upward-diagonalization} is an antecedent
of our proof approach.

The notion of ``robustly
strong'' is made up of  two  components---one stating
that for all sets
and all inputs the reducing machine 
has at least one non-{\bf{}?}~path,
and the other stating that for all sets and all inputs the
reducing machine does not simultaneously have {\bf acc}
and {\bf rej} paths.  
Each component has been separately studied before
in the literature, in
different contexts (see Section~\ref{s:relations}).  
By considering each of these two requirements in
conjunction with strong reductions, we obtain two natural new
reductions whose power falls between that of strong reductions
and that of robustly strong reductions.
Section~\ref{s:relations} studies the relative power of Turing
reductions, of strong reductions, of robustly strong reductions, and
of our two new reductions.  In some cases we prove absolute
separations.  In other cases, we see that the relative computation
power is tied to the $\pisnp$ question.  Curiously, the two new
reductions are deeply asymmetric in terms of what is currently
provable about their properties.  For one of the new
reductions, we show that if it differs
from Turing reductions then $\pisnotnp$.  For the other, we prove that
the reduction does differ from Turing reductions.

In Section 5, we  discuss some issues regarding what
collapses of the polynomial time hierarchy occur if sparse sets exist
that are hard or complete for $\np$ with respect to the new
reductions.  One of the new reductions extends the reach of 
hardness results.

\section{Two New Reducibilities}
For each NPTM $N$ and each set $D \subseteq
\sigmastar$, define
$out_{N^D}(x) 
= \{ y \condition y\in \{ \mbox{\bf{}acc},\mbox{\bf{}rej},
\mbox{\bf{}?} \} 
\land$ some computation path of $N^D(x)$ has outcome $y\}$.
As is standard, for each nondeterministic machine $N$
and each set $D \subseteq \sigmastar$, let $L(N^D)$ 
denote the set of all $x$ for which $\mbox{\bf{}acc}\in out_{N^D}(x)$.
For each nondeterministic machine $N$
and each set $D\subseteq\sigmastar$,
let $\lrej(N^D)$ denote the set
of all $x$ for which $\mbox{\bf rej}\in out_{N^D}(x)$.
A computation $N^D(x)$ is called 
{\it underproductive} if $\{ \mbox{\bf acc},\mbox{\bf rej} \}
\not\subseteq out_{N^D}(x)$.  That is, $N^D(x)$ does not 
have as outcomes both {\bf acc} and {\bf rej}.
$N^D$ is said to be
{\em underproductive\/} if, 
for each string~$x$,
$N^D(x)$ is underproductive.
That is,
$L(N^D) \cap \lrej(N^D) = \emptyset$.
Underproductive 
machines were introduced by
Buntrock~\cite{bun:thesis:log}.
Allender et al.~\cite{all-bei-her-hom:j:nondet-ae-hier}
have shown underproductivity to be very useful
in the study of almost-everywhere complexity
hierarchies for nondeterministic time
classes.

\begin{figure}[!tp]
\begin{center}
\vspace*{0.6in}
\setlength{\unitlength}{0.25mm}
\begin{picture}(125,125)(-50,-50)
\linethickness{0.25mm}
\qbezier(0,50)(25,25)(50,0)
\qbezier(50,0)(75,25)(100,50)
\qbezier(100,50)(75,75)(50,100)
\qbezier(50,100)(25,75)(0,50)
\qbezier(50,0)(50,-13)(50,-45)
\put(-15,40){\makebox(0,0)[b]{\boldured}}
\put(75,-5){\makebox(0,0)[b]{\boldrsred}}
\put(115,40){\makebox(0,0)[b]{\boldored}}
\put(75,100){\makebox(0,0)[b]{\boldsnred}}
\put(75,-45){\makebox(0,0)[b]{\boldturingred}}
\end{picture}
\end{center}
\caption{\label{f:diamond}Inclusions of
Proposition~\ref{p:inc}}
\end{figure}
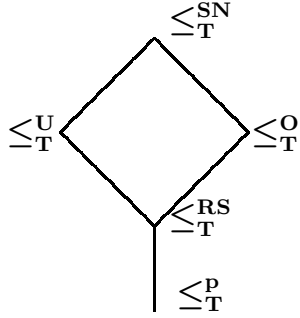

A computation $N^D(x)$ is called {\it overproductive} if 
$out_{N^D}(x)\neq 
\{\mbox{\bf ?}\}$.
A machine $N^D$ is said to be
{\em overproductive\/} if, for each 
string~$x$, $N^D(x)$ is overproductive.  
Equivalently, $L(N^D) \cup \lrej(N^D) = \sigmastar$.

We say that $N$ is {\em robustly overproductive\/}
if for each $D \subseteq \sigmastar$ it holds
that $N^D$ is overproductive.
We say that $N$ is {\em robustly underproductive\/}
if for each $D \subseteq \sigmastar$ it holds
that $N^D$ is underproductive.

Using underproductivity, overproductivity, and robustness, we may
now define strong and robustly strong reductions, which have been
previously studied.  We also introduce two intermediate reductions,
obtained by limiting the robustness to just the overproductivity or
the underproductivity.\footnote{{}\protect\singlespacing{}The
literature contains various notations for strong reductions (also
known as strong nondeterministic reductions).  We adopt the notation
of Long's paper~\protect\cite{lon:j:nondeterministic-reducibilities},
i.e., $\snred\!$.  However, we note that some papers use other notations,
such as $\leq^{\rm SN}$, $\leq_{\rm T}^{\rm sn}$, and $\leq_{\rm
T}^{\rm p,\np \cap
\conp}$.  For the three other reductions we discuss, we replace the SN with a
mnemonic abbreviation.  For robustly strong we follow Gavald\`{a} and
Balc\'{a}zar~\protect\cite{bal-gav:j:rob} and use RS\@.  For brevity,
we use O as our abbreviation for our ``strong and robustly
overproductive'' reductions, and we use U as our abbreviation for our
``strong and robustly underproductive'' reductions.}

\begin{definition}\label{def:twored}~
\begin{enumerate}
\item {\cite{lon:j:nondeterministic-reducibilities}, 
see also \cite{sel:j:enumeration-reducibility} }
(``strong reductions'') \quad
$A \snred B$ 
if there is an NPTM $N$ such that 
$N^B$ is overproductive, $N^B$ is underproductive, 
and $A=L(N^B)$.
\item { \cite{bal-gav:j:rob}}
(``robustly strong reductions'') \quad
$A \rsred B$ if $A \snred B$ via an NPTM $N$, and  
$N$ is both robustly overproductive and
robustly underproductive.

\item 
(``strong and robustly underproductive reductions'' or,
for short, ``U-reductions'') \quad
$A \ured B$ if $A \snred B$ via an NPTM $N$ that is 
robustly underproductive.
\item 
(``strong and robustly overproductive reductions'' or,
for short, ``O-reductions'') \quad
$A \ored B$ if $A \snred B$ via an NPTM $N$ that is
robustly overproductive.
\end{enumerate}
\end{definition}

The trivial containment relationships are shown in
Proposition~\ref{p:inc} and Figure~\ref{f:diamond}.  
In 
this paper we ask whether some edges of the diamond pictured
in Figure~\ref{f:diamond} might collapse, and in particular we seek
necessary conditions and sufficient conditions for such collapses.
\begin{notation}
For each well-defined reduction $\leq_a^b$,
let \mbox{\boldmath{$\leq_a^b$}} denote
$\{ (A,B) \condition A \leq_a^b B \}$.
\end{notation}
\begin{proposition}\label{p:inc}
$\boldturingred \subseteq \boldrsred
{  
{\subseteq \boldured \subseteq}
\atop
{\subseteq \boldored \subseteq}
}
\boldsnred\!$.
\end{proposition}

Using different
terminology, robust underproductivity 
(though not $\ured\!$)
has been introduced into
the literature by Beigel~(\cite{bei:c:up1}, see
also~\cite{har-hem:j:rob}), and the following theorem will be of use
in the present paper.
\begin{theorem}{}\label{t:rueas}
{ (\cite{bei:c:up1}, see also~\cite{har-hem:j:rob})}
\quad
If NPTM $N$ is robustly underproductive, then $$(\forall A) (\exists L
\in \p^{{\rm SAT}  \oplus A}) [\lrej(N^A) \subseteq L \subseteq
\overline{ L(N^A)}].$$
\end{theorem}

Theorem~\ref{t:rueas} says that 
if a machine is robustly underproductive, then for every
oracle there is a relatively simple set that separates its acceptance
set from its $\lrej$ set.  In particular, if $\pisnp$ and $N$ is a
robustly underproductive machine, then for every oracle $A$ it holds
that $L(N^A)$ and $\lrej(N^A)$ are $\p^A$-separable.

As is standard, we say that a set $S$ is {\em sparse\/} if 
there is a polynomial $r$ such that, for each $n$,
$||S^{\leq n}|| \leq r(n)$.
Using
different terminology, ``robust with respect to sparse
sets''-overproductivity (though not $\ored\!$) has 
been introduced into the literature
by Hartmanis and Hemachandra~\cite{har-hem:j:rob}, and the following
theorem will be of use in the present paper. 
\begin{theorem}{}\label{t:hh-over}
{ \cite{har-hem:j:rob}}
\quad
If NPTM $N$ is such that for each sparse set $S$ it holds that $N^S$
is overproductive, then for every sparse set $S$ there exists a 
predicate $b$ computable in $\fp^{{\rm SAT} \oplus S}$ such that, for all
$x$, $\{x \condition b(x)\}
\subseteq L(N^S)$ and $\{x \condition  \neg b(x)\}
\subseteq \lrej(N^S)$, where $\fp$ denotes the 
polynomial-time computable functions.
\end{theorem}

Theorem~\ref{t:hh-over} says 
that if a machine is ``robustly with respect to sparse
oracles''-overproductive, then for every sparse oracle there is a
relatively simple function that for each input correctly declares
either that the machine has accepting paths or that the machine has
rejecting paths.  Crescenzi and Silvestri~\cite{cre-sil:c:sperner}
show via Sperner's Lemma that Theorem~\ref{t:hh-over} fails when the
sparseness condition is removed, and their proof approach will
be of use in this paper.

It is known that SN reductions and RS reductions 
have nonuniform characterizations.  In particular,
for every reducibility $\leq_a^b$ and every class $\calc$, let
$$\red{a}{b}{\calc} = \{ A \condition
(\exists B \in \calc)[A \leq_a^b B]\}.$$
Gavald\`{a} and Balc\'{a}zar 
proved the following result.
\begin {theorem}\label{spcl} 
{  \cite{bal-gav:j:rob}}\quad
\begin{enumerate}
\item $\red{\rm T}{\rm SN}{\sparse} = 
\np/\poly \inter \conp/\poly$.
\item \label{p:rs}
$\red{\rm T}{\rm RS}{\sparse} = (\np \inter \conp)/\poly $.
\end{enumerate}
\end{theorem}

We note
in passing that the downward closures of the sparse sets under our two
new reductions have analogous characterizations, albeit somewhat
stilted ones.
We say $A\in \np/\poly \inter \conp/\poly$ via the pair $(M,N)$ of NPTMs if
there is a sparse set $S$ such 
that $A=L(M^{S})$ and $A = \overline{L(N^{S})}$.
%
%
Hartmanis and 
Hemachandra~\cite{har-hem:j:rob} 
defined {\it robustly $\sigmastar$-spanning pairs of machines} $(M,N)$ 
to be pairs having
the property $L(M^{X})\cup L(N^{X})=\sigmastar$ for every oracle $X$, and
{\it robustly disjoint pairs} to be pairs having 
the property 
$L(M^{X})\cap L(N^{X})=\emptyset$ for every oracle $X$.
Using these notions  
we note the following characterizations.
$A\in \red{\rm T}{\rm O}{\sparse}$ if 
and only if $A\in \np/\poly \inter \conp/\poly$ 
via some robustly $\sigmastar$-spanning  pair $(M,N)$ of NPTMs.
$A\in \red{\rm T}{\rm U}{\sparse}$ if and 
only if $A\in \np/\poly \inter \conp/\poly$ 
via some robustly disjoint  pair $(M,N)$ of NPTMs.

\section{A Strong Separation of $\snred$ and $\rsred$}\label{sec:xy}

It follows from each 
of Section~\ref{s:relations}'s
Theorems~\ref{snod} and~\ref{rsod}, both of 
which have relatively simple proofs, that the reducibilities
$\snred$ and $\rsred$ are distinct.  However, 
more can be said.  The separation of these two reductions turns out to
be extremely strong, namely, for every recursive set $A\not\in \npinterconp$,
there exists a recursive set $B$ such that 
$A$ is strongly reducible to $B$ but
$A$ is not robustly strong reducible to $B$.  This is  Theorem~\ref{rec}. 
As noted in Section~\ref{sec:intro},
this claim cannot be generalized to include $\np \cap \conp$ since
$\np \cap \conp$ is the zero degree of $\rsred\!$, as has been pointed out 
by
Gavald\`{a} and Balc\'{a}zar~\cite{bal-gav:j:rob}.
Theorem~\ref{rec}
was first 
stated in 
Gavald\`{a} and Balc\'{a}zar's 1991 paper~\cite{bal-gav:j:rob}.
The diagonalization proof
given there correctly establishes $A\notrsred B$, but it fails to establish
$A\snred B$. The main error is the following: In the proof there is a
passage~\cite[p.~6, lines $21$--$25$]{bal-gav:j:rob}
where 
a certain 
word $x$ is searched for.
If such an $x$ is
found,
then $B$ is augmented by some suitably chosen word (triple). Now it is true
that such an $x$ must always exist.   However, 
it might be huge, and then between
this $x$ and the previous one, say $x'$, no coding has been done, i.e., for
all
$z$ between $x'$ and $x$, no triple $\langle z,y,0\rangle$ or
$\langle z,y,1\rangle$ with $|z| = |y|$ has been added to $B$.
Thus, the condition ``(i)'' of \cite[p.~6]{bal-gav:j:rob},
which is intended to guarantee $A\snred B$, is violated. 

We now turn towards the proof 
of Theorem~\ref{rec}.  However,
we first prove  the following claim.
\begin{theorem}
\label{notrec}
$(\forall \mbox{\rm{} recursive } A\not\in \npinterconp)(\exists B)[A \snred B 
\wedge A \notrsred B]$.
\end{theorem}
\sproof
We distinguish two cases:
{\it Case 1:} $A \in \et$ and
{\it Case 2:} $A \not\in \et$.
In the first case, we will show that if no such $B$ exists, then
$A \in \npinterconp$.  In the second case, we will show that
if no such $B$ exists for our $A$, then in fact $A \in \et$, thus
generating a contradiction to $A\not\in \et$.

\noindent
{\it Case 1:} \quad
Suppose $A \in \dtime[2^{n^k}]$ for some constant $k$.
Our set $B$, as finally constructed, 
will satisfy the following condition.

\begin{cond}\label{main-cond}
For each $x \in \sigmastar$,
\begin{enumerate}
\item $x \in A \longrightarrow   
\left(
(  \exists y) [ |y| = |x|^k \land  \langle x, y, 1 \rangle \in B ]
 \wedge   (  \not{\!\exists} y')[ |y'| = |x|^k \land
   \langle x, y', 2 \rangle \in B ]
\right)$, and
\item $x \not\in A \longrightarrow   
\left(
(  \not{\!\exists} y)[ |y| = |x|^k \land \langle x, y, 1 \rangle \in B ]
                \wedge   (  \exists y')[ |y'| = |x|^k \land
      \langle x, y', 2 \rangle \in B ]
\right)$.
\end{enumerate}
\end{cond}

A subset of $\sigmastar$ is identified with its characteristic sequence 
according to  lexicographic order. 
The {\it $n$-segment}, $B^{<n}$, of a set $B$ is defined 
to be
the initial segment of this sequence including all words of
length less than $n$.

A set $B$ satisfying Condition~\ref{main-cond} 
is called {\it admissible}.
Each initial segment of an admissible set is also called admissible. A 
 set $C$
is called  a {\it consistent extension} of an admissible initial 
$n$-segment 
$I$ if $C$ is admissible and $C^{<n}=I$.

For each $B$ satisfying 
Condition~\ref{main-cond}, it clearly holds that 
$A \snred B$.
We will define $B$ in such a way that no  NP machine with three
final states---{\bf{}acc}, {\bf rej}, and~{\bf{}?}---can  robustly 
strong-reduce $A$ to $B$.
Assume a list $\widehat{L}$ of 
\mbox{{\bf{}acc}/{\bf{}rej}/{\bf{}?}-final-state} NPTMs
created by pairing every 
\mbox{{\bf{}acc}/{\bf{}rej}/{\bf{}?}-final-state} nondeterministic Turing
machine
with every 
clock of the form $n^{\widehat{k}}+{\widehat{k}}$, $\widehat{k}>0$.  
Note that
each $n^{\widehat{k}}+{\widehat{k}}$ is a monotonically
increasing function on the natural numbers.
Initially,
$i=1$, $n=0$, and $B^{<0} = \emptyset$.
Note that this is admissible.
Suppose now $i \geq 1$, and
 that $B$ has been determined up to  $B^{<n}$, and that
$B^{<n}$ is admissible.

\noindent
{\it Stage i} \quad Let $N$ be the $i$th machine on the list
$\widehat{L}$.
Let $p$ be the polynomial of the (clearly attached) clock 
that (for all oracles) upper-bounds the 
running time of $N$.  Define 
$n_{0}$
to be the smallest number $m>n$ such that:
\begin{equation} \label{e:b}
(\forall b\in\{1,2\}) (\forall x,y \in \sigmastar)
[ \left( | \langle x,y,b\rangle | \geq m \land |y| = |x|^k \right)
\longrightarrow p(|x|) < 2^{|x|^k}].
\end{equation}
Extend
$B^{<n}$ to an admissible segment $B^{<n_{0}}$.
(This can always be done, as the very fact that $B^{<n}$
is an admissible $n$-segment implies that there is an 
admissible set $B$ of which it is a prefix, so the $n_0$-segment
of that set $B$ is an extension of $B^{<n}$ and must also
be admissible.)

Now we consider the following question.
(Recall that we have defined above the notion 
of a consistent extension of an admissible initial segment.)
\begin{question}\label{main-question}
Is there a consistent extension $C$ of $B^{<n_{0}}$ such that
either $(\exists x \in A)[N^{C}(x)$ has a rejecting computation$]$
or $(\exists x \not\in A)[N^{C}(x)$ has an accepting computation$]$?
\end{question}

If the answer to Question~\ref{main-question} 
is ``yes,'' then we fix such an extension $C$ 
for the lexicographically smallest applicable $x$, determine $m$ by
$m=\max\{n_{0}+1,p(|x|)\}$,  and extend $B^{<n_{0}}$ to $B^{<m}$,
which is chosen to be the $m$-segment of $C$. This choice of the extension 
preserves
all rejecting paths (if the first case 
of Question~\ref{main-question} occurs) and all accepting paths 
(if the
second case of Question~\ref{main-question} occurs).
If the answer to Question~\ref{main-question}
is ``no,'' then nothing further will be done in Stage~$i$.
Notice that we do not claim that the answer to
Question~\ref{main-question} is
recursively computable. However, this is not a
problem, as 
Theorem~\ref{notrec} does not require 
$B$ to be recursive.

Clearly, Question~\ref{main-question}
is answered ``yes'' infinitely often.
($\widehat{L}$, as any reasonable enumeration of Turing machines
does, contains infinitely many machines $N$ such that
for all oracles $C$, $L(N^C) \not \subseteq A$ or $L_{rej}(N^C) 
\not \subseteq \overline{A}$.)
Since Question~\ref{main-question}
is answered ``yes'' infinitely often,
clearly the length to
which $B$ is constructed increases unboundedly.
Thus $B$, as finally 
constructed,
is admissible (and not a finite initial segment).
Since $B$ is admissible,
it follows that $A$ is strongly reducible to $B$.

It
remains to show that $A$ is not robustly strong reducible to $B$. We prove 
this by contradiction.  Suppose $A\rsred B$ via some machine $N$
from list $\widehat{L}$.  (If $A\rsred B$ via some machine not on
list $\widehat{L}$ then certainly $A\rsred B$ via some machine 
on $\widehat{L}$, so the assumption that $N$ is chosen from
$\widehat{L}$ can be made without loss of generality.)
Let $p$ be 
the
polynomial enforced by the 
clock of machine $N$;  recall that this is a 
monotonically increasing polynomial that, 
independent of the oracle, upper-bounds the running time 
of $N$.  Assume that $N$ was considered in Stage~$i$ 
 and let $B^{<n_{0}}$ be the initial segment of $B$ that 
was constructed at the beginning of Stage $i$ as admissibly extended
as described right after Equation~\ref{e:b}.
We prove the following claim.

\noindent
{\it Claim} \quad {\it If  $A \rsred B$
(where $B$ is as just constructed), then  $A \in \npinterconp$.}

We will describe an NPTM $M$ that witnesses the membership
$A \in \npinterconp$.  The machine $M$ remembers the finite
initial segment $B^{<n_{0}}$.  On each input $x$,
$M$ will simulate without an oracle
the computation of $N$ on $x$ using a certain
oracle that actually depends on~$x$.

For each $x$, let the integer $m_x$ be defined by
$$ m_{x}=\mbox{max}\{n \condition 
2^{n^{k}}\leq p(|x|)\},$$
and let 
\begin{eqnarray*}
B_{x}& = & B^{<n_0} \,\cup\,
\{  \langle x', 0^{|x'|^k}, 1 \rangle \condition
 |\langle x', 0^{|x'|^k}, 1 \rangle | \geq n_0 \wedge |x'| \leq m_x \wedge
x' \in A \}\\
 &  &
 \quad \quad \, \, \, \cup \, 
\{  \langle x', 0^{|x'|^k}, 2 \rangle \condition
 |\langle x', 0^{|x'|^k}, 2 \rangle | \geq n_0 \wedge |x'| \leq m_x \wedge
x' \not\in A \}.
\end{eqnarray*}

Now we describe $M$.
On input $x$, $M$
 simulates $N^{B_{x}}(x)$, where queries are handled as 
follows:
Short queries (i.e., of length~$< n_{0}$) are answered 
directly by the finite set
$B^{<n_{0}}$. Each other query is answered ``no,'' unless
it is  of the form
$\langle x', 0^{|x'|^k}, 1 \rangle$ or  $\langle x', 0^{|x'|^k}, 2 
\rangle$, where $|x'| \leq m_x$. For each such query,
use  the $\et$ algorithm
for $A$ to determine membership of $x'$ in $A$.  This algorithm runs
in time at most $2^{|x'|^k}  \leq p(|x|)$, since $|x'| \leq m_x$.  
If $x' \in A$ then answer ``yes'' to
$\langle x', 0^{|x'|^k}, 1 \rangle$  
and answer ``no'' to $\langle x', 0^{|x'|^k}, 2 \rangle$.
If $x' \not\in A$ then do exactly the opposite.

This shows:
\begin{quote}
$M$ runs in nondeterministic polynomial time
and, for each $x$, simulates the work of $N^{B_{x}}(x)$.
\end{quote}
Since we assumed that $N$ is robustly strong, 
it follows that, for all $x$ and for all
oracle sets, $N$ on $x$ using the oracle yields either some accepting
computation and no rejecting computation, or some rejecting computation
but no accepting computation.  
This is true even though the oracle $B_{x}$
used by $N$  depends on $N$'s input, $x$. 
Thus the only issue is whether the decision made
along each accepting or rejecting path is correct.

Suppose some accepting or rejecting path is incorrect. Thus, for some $x$,
either $x \in A$ and yet some path as described above rejects,
or $x \not\in A$ and yet some path as described above accepts.
However, for this particular $x$ and a particular such path
we can extend  $B^{<n_0}$ 
(in fact, can extend  ${B}_{x}$ in light of the comments
later in this paragraph regarding the pairing function) 
consistently in such a way that
such a rejecting or accepting computation path is preserved.
The key point is that the number of queries is at most $p(|x|)$, which
is strictly less than $2^{|\widetilde{x}|^k}$,
the number of available $y$'s for any $\widetilde{x}$ with 
$|\widetilde{x}| > m_x$.
To claim that all these $y$'s are actually available, we note
that we are assuming certain properties of the pairing function, 
namely, that if $|a| = |a'|$, $|b|=|b'|$, 
$c\in\{1,2\}$, and $c' \in \{1,2\}$, then 
$|\langle a,b,c\rangle| = |\langle a',b',c' \rangle|$.  We also 
assume, as is standard, that if one increases the length of any
one input to the pairing function the length of the output does
not decrease.  We do not require that the pairing function be
onto, though we do require it be 1-to-1.
Thus, a consistent extension of ${B}_{x}$ can be found that
preserves the particular path.

However, if that is the case, then the answer to the initial question,
Question~\ref{main-question},
during the construction of $B$ must have been ``yes,'' and the construction
of $B$ would have explicitly ruled out 
the possibility that the machine $N$ accepts $A$
with oracle  $B$.  This contradicts our supposition 
regarding there being some incorrect path.
Thus, it must be the case that for all $x$ the described
computation of $M$ is such that
(a)~if $x \in A$ then some path accepts
and no path rejects, and
(b)~if $x \not\in A$ then some path rejects
and no path accepts.
This is a proof that 
$A \in \npinterconp$.  Thus our assumption (for the 
current case---$A\in \et$) that the  $B$ constructed does not
 satisfy
Theorem~\ref{notrec} leads to a contradiction,
as in fact $A\not\in \np\inter\conp$.  This concludes the 
proof of 
Case~1 ($A \in \et$).


\medskip 

\noindent
{\it Case 2:} 
Suppose $A \not\in \et$. The proof structure is as follows: 
Similarly to Case 1 we construct a set $B$ such that $A\snred B$. 
However,
if  $A\rsred B$ then we show that this implies that  $A \in \et$,
yielding a contradiction to $A\not\in \et$.

We require that $B$ satisfy 
Condition~\ref{main-cond}
for $k=1$:
\[
x\in A \longrightarrow 
\mbox{\large$($}
(\exists y)[|y| = |x|  \wedge
\langle x,y,1\rangle \in B]
\wedge (\not{\!\exists} z)[|z|=|x|  \wedge 
\langle x,z,2\rangle \in B]
\mbox{\large$)$}, \mbox{\rm and }
\]
 \[x \not\in A \longrightarrow
\mbox{\large$($}
 (\not{\!\exists} y)[|y| = |x|  \wedge 
\langle x,y,1\rangle \in B]
\wedge (\exists z)[|z|=|x| \wedge \langle x,z,2\rangle
 \in B]
\mbox{\large$)$}.
\]
The construction of $B$ is as in Case 1. Thus we get an admissible $B$, and
hence $A\snred B$.

The crucial claim is the following:

\noindent {\it Claim:}
{\it If $A \rsred B$ then $A \in \et$.}

Let us assume $A \rsred B$ via $N$. 
For reasons analogous to those discussed 
in {Case~1}, we may without loss of
generality assume that $N$ is chosen 
from the list $\widehat{L}$.  Let $p$ be the
(nondecreasing) polynomial clock upper-bounding,
independent of the oracle, the running time of $N$.
Assume that $N$ was considered in Stage~$i$ 
and let $B^{<n_{0}}$ be the initial segment of $B$ that
has been constructed  at the beginning of  Stage~$i$
and admissibly extended
as described right after Equation~\ref{e:b}.

For each $x$, let the integer $m_x$ be defined by 
$$ m_{x}=\mbox{max}\{n \condition 
2^{n}\leq p(|x|)\}.$$
Note that $m_x = \bigo(\log |x|)$, and an upper-bound on
the constant 
of the $\bigo$ can be seen immediately given the 
polynomial $p$.  
For each $x$, let 
\begin{eqnarray*}
B_{x}& = & B^{<n_0} \, \cup \,
\{  \langle x', 0^{|x'|}, 1 \rangle \condition
 |\langle x', 0^{|x'|}, 1 \rangle | \geq n_0 \wedge |x'| \leq m_x \wedge
x' \in A \}\\
 &  &
 \quad \quad \, \, \, \cup \,
\{  \langle x', 0^{|x'|}, 2 \rangle \condition
 |\langle x', 0^{|x'|}, 2 \rangle | \geq n_0 \wedge |x'| \leq m_x \wedge
x' \not\in A \}.
\end{eqnarray*}
We will describe a deterministic computation that 
for any given $x$
simulates
$N^{B_x}(x)$. 
For a given $x$, 
the algorithm either will use table lookup or 
will examine $2^{p(|x|)}$ nondeterministic 
paths of
$N(x)$ in turn.

Since $m_x = \bigo(\log |x|)$, there are at most a finite
number of inputs $x$ for which $|x| \leq m_x$.
On input $x$, if $|x| \leq m_x$ then use finite table lookup
to determine the correct answer.
Now suppose $|x| > m_x$.

Let $p'(n) = p(n)+ n^2$. Then $p'(n) \geq p(n)$, $p'(n) \geq n^2$ and,
like $p(n)$, $p'(n)$
is  monotonic increasing.
We will inductively assume that for all $x'$ such that $|x'| < |x|$,
 ``$x' \in A$?'' can be decided in time 
$2^{p'(|x'|)}$.  Upon input $x$,  $|x|=n$, we first compute 
$\chi_A(x')$ for all $|x'| < |x|$.
This takes time at most $\sum_{i=0}^{n-1} 2^i 2^{p'(i)}$,
which is bounded by $2^{p'(n-1) + n}$, by the fact that
$p$ is monotonic and thus this is a geometric series.
Next we write down $p(n)$ bits, as nondeterministic
moves of $N(x)$.  We will cycle through all such $ p(n)$ bits.
For a particular sequence of $p(n)$ bits, we simulate 
the computation of $N$ on $x$ as follows:
Whenever $N(x)$  makes a query $q$ of length less than $ n_0$,
we answer it according to $B^{<n_0}$.
If a query $q$ has length $|q| \geq n_0$, and is of the relevant form
$\langle x', 0^{|x'|}, 1 \rangle$ or
$\langle x', 0^{|x'|}, 2 \rangle$, and $|x'| \leq m_x$
(note that this implies
that $|x'| \leq m_x < |x|$),
then we answer according to $\chi_A(x')$.
For all other queries, the answer is ``no.''
Thus, the computation for $x$ with $|x|=n$
takes time at most 
\[2^{p'(n-1) + n} + 2^{p(n)} p(n).\]
We have $p(n) < 2^n$ since $n = |x| >  m_x$.
It is easy to verify that $p(n) + n \leq p'(n) -1$
as well as $p'(n-1) + n \leq p'(n) -1$,
by the definition of $p'(n)$, and the
monotonicity of $p(n)$. Thus the time taken to
simulate  the computation of $N$ on $x$ is at most
$2^{p'(n)}$, completing the induction.
This completes the proof of the {\it Claim}, namely $A \in \et$,
and thus we reach a contradiction in~Case~2.

The proof of Theorem~\ref{notrec} is complete.~\qed

Now we prove the main claim of this section.
\begin{theorem} \label{rec}
$(\forall \mbox{\rm{}  recursive }
A\not\in \npinterconp)(\exists \mbox{\rm{}   recursive } B)
[A \snred B \wedge A \notrsred B]$.
\end{theorem}
\sproof
For an initial $n$-segment $I$ we define $X*I=(X\setminus X^{<n}) \cup I$.
We make the following three preliminary claims.
As they are clear, we state them without proof.

\noindent {\it Claim 1:} {\it For each initial segment $I$:
If $A\snred B$ and $C=B*I$, then $A\snred C$.}

\noindent 
{\it Claim 2:} {\it Let $A$
and $B$ be chosen according to Theorem \ref{notrec} and
let $C=B*I$ for some initial segment $I$. Then every machine $N$ strongly
reducing $A$ to $C$ must fail to be strong for some extension $E$ of $I$.}

\noindent {\it Claim 3:} {\it Every
admissible initial segment $I$ has a consistent
extension $C$ such 
that $A\snred C$, but no machine reducing $A$ strongly
to $C$ is strong for all
extensions of~$I$.}

Now we prove Theorem~\ref{rec}. We modify the construction of $B$. Let
us assume that at 
the beginning of Stage $i$ the admissible initial segment $I$
 is
available as the result of the previous steps. Let the machine $N$ be
considered in Stage~$i$.

We check by a systematic search whether
\begin{itemize}
\item[(1)]
there exists 
a finite initial segment that is an
admissible extension of $I$ that 
 witnesses that $A$ is not correctly reduced to this extension by $N$
(recall that the definition of an admissible initial segment means
each such initial segment is an initial segment of 
some admissible set),
or
\item[(2)] 
there exists a finite initial segment that is an extension
of $I$ that witnesses that $N$ fails to be 
strong using this
extension as its oracle.~(Here we can allow even
inadmissible extensions.)
\end{itemize}

Call these two cases type-1 and type-2.
If a type-1 extension
is found first, then extend $B$ in an admissible way to preserve a
computation witnessing a contradictory computation (i.e., an accepting path
if $x \not\in A$, and a rejecting path if $x\in A$). If 
a type-2 extension is found first,
then nothing is done in this stage.

\noindent {\it Claim:} {\it One of the two cases must happen.}

Let $N$ be strong for all extensions of $I$. Since $I$ is admissible, Claim 3
yields an admissible $C$ such that $A\snred C$, but no machine strongly
reducing $A$ to $C$
can be  robustly strong for all extensions of $I$. Since
$N$ is strong for all extensions of $I$,
from Claim~3 we can conclude
that $N$ does not strong-reduce $A$ to $C$.
So
$N$ does not reduce $A$ to some 
particular finite initial segment of 
$C$ (one long enough to witness the non-reduction of $A$ to 
$C$ via $N$).
Thus a type-1 extension will be
found unless a type-2 extension is encountered sooner.

Suppose now 
that $N$ is not strong for some extension of~$I$. Then this will
become apparent at some {\it finite} length for some (not necessarily
consistent) finite
extension of $I$ on some input. Hence 
a type-2 extension is found unless a type-1 extension is found 
sooner.
This modification of the construction of $B$ shows that $B$ is the union of
a growing sequence of effectively constructed
finite initial segments. Thus, $B$ is recursive.~\qed

More generally, 
the proof actually shows that a $B$ recursive in $A$ can be found
to satisfy the theorem.

One can ask whether the difference of $\snred$ and $\rsred$ is so strong
that
the following statement holds:
$(\forall \mbox{\rm{}  recursive } 
B \not\in \npinterconp)(\exists \mbox{\rm{} recursive } A)
[A\snred B \wedge A\notrsred B]$.
This can be reformulated in terms of reducibility downward closures:
$(\forall \mbox{\rm{}  recursive } 
B\not\in \npinterconp) [\red{{\rm T}}{\rm RS}{B}
\subsetproper
\red{\rm T}{\rm SN}{B}]$.
However, this claim
is false. 
Intuitively, if $B$ is chosen to be of appropriately great
structural richness,
the
differences between the two reductions may be too fine to still be
distinguishable in the presence of $B$.  For instance, if $B$ is
an
${\rm EXPSPACE}$-complete set, and thus is certainly not contained
in
$\npinterconp$,
then
for every $A \in \red{\rm T}{\rm SN}{B} ={\rm NP}^{B} \cap {\rm coNP}^{B}
=
{\rm EXPSPACE}$ we have $A\leq^{{\rm p}}_{{\rm m}} B$ and hence $A\rsred
B$,
i.e.,
$\red{\rm T}{\rm SN}{B}=\red{\rm T}{\rm RS}{B}$.

\section{Comparing the Power of the Reductions}\label{s:relations}
Long~\cite{lon:j:nondeterministic-reducibilities} 
proved 
that strong and 
Turing 
polynomial-time reductions differ. 
More precisely, he proved the following result.
\begin{theorem}\label{long}
{} \cite{lon:j:nondeterministic-reducibilities}  \quad
$(\forall \mbox{\rm{}  recursive } A\not\in {\rm P})
(\exists \mbox{\rm{}  recursive } B)
[A\snred B \wedge A\not\leq_{\rm T}^{\rm p}\,
B]$.
\end{theorem}

Consequently, at least one of the edges in Figure 1 must represent a strict
inclusion. Indeed, we can show that strong reductions differ from both overproductive
and underproductive reductions. 

\begin{theorem}{}\label{snod}~
\begin {enumerate}
\item \label{p:overpart}
$(\exists \mbox{\rm{}  recursive } A)
(\exists \mbox{\rm{}  recursive } B)
[A \snred B \wedge A \notored B].$ 
Indeed, we may even achieve this via a recursive sparse set~$B$
and a recursive tally set~$A$.

\item \label{p:underpart}
$(\exists \mbox{\rm{} recursive } A)
(\exists \mbox{\rm{}  recursive } B)
[A \snred B \wedge A \notured B].$
Indeed, we may even achieve this via a recursive sparse set~$B$
and a recursive tally set~$A$.
\end{enumerate}
\end{theorem}
\sproof
For a given set $B$ define $A$ (implicitly, $A_B$) by
\begin{equation}
\label{1}
A=\{0^{i}\condition (\exists y)[| y | =i \wedge \langle 0^{i},y,1\rangle
\in B]\}.
\end{equation}
$B$ is constructed by diagonalization, and we make sure that
\begin{equation}
\label{2}~
(\forall i)(\exists \mbox { exactly one}\hspace{0.5 em}u\in\{0,1\})(\exists
\mbox{ exactly one}\hspace{0.5 em} y)
[ \langle 0^{i},y,u\rangle \in B \wedge | y | =i ].
\end{equation}
Assume that $B$ has the property 
of Equation~\ref{2} and that $A$ is defined by 
Equation~\ref{1}. Then $A$
strongly reduces to $B$ via a machine $N$ that is described as 
follows (this is simply making explicit what is 
implicitly clear from Equation~\ref{1}):
\begin{quote}
An input $x$ is rejected if it is not an element
of $\{0\}^{*}$. Otherwise, $x$ is  
an element
of $\{0\}^{*}$, say $x=0^i$.
Then $N$ 
nondeterministically generates all $y$ such that $|y| =i$, and using the
oracle $B$, for each $y$ and $u \in \{0,1\}$, it finds out whether
$\langle 0^{i},y,u\rangle \in B. $ A path accepts if on this path
$\langle 0^{i},y,1\rangle \in B $ is determined, and it rejects if
$\langle 0^{i},y,0\rangle \in B$ is determined. In all remaining cases, the
path is a no-comment
(i.e.,~{\bf{}?})~path. 
\end{quote}
By Equation~\ref{2}, on each input $N$ reaches on exactly 
one path the correct answer to the question of whether $x\in A$, and all remaining paths are no-comment paths.

For the construction of $B$ we assume an effective enumeration,
$E$, of all
nondeterministic polynomial-time oracle machines. $B$ is constructed
to be the union $B=\cup_{i=0}^{\infty} B_{i}$, where $B_{i}$ is determined
(in a recursively computable way)
in Stage $i$ of the construction, and $B_0\subseteq
B_{1}\subseteq B_{2}\subseteq \ldots$
 are finite initial segments of $B$. This ensures 
that $B$ is recursive,  which 
ensures that $A$ also is recursive.  Set $B_0 = \emptyset$ and 
let $i$ initially equal~1.  As a formal choice to avoid problems
in Stage~1, we act as if the ``$n$'' of Stage~0 were $-1$.

\smallskip

\noindent {\it Start of stage construction}\nopagebreak

\nopagebreak\noindent
{\it Stage $i$}.  Let $M$ be the $i$th machine in 
our enumeration~$E$\@.  
Let the 
polynomial $p$ be a bound on the computation time of $M$. Choose $n$ 
sufficiently large (so large that the lengths of all queries occurring in all
previous stages are less 
than $n$ and that $p(n)<2^{n}$). 
To 
help maintain Equation~\ref{2} we will do some coding now.
For each $\widetilde{n}$ that is strictly less than the $n$ we just 
set, but strictly greater than the $n$ that was set in Stage~$i-1$,
put into $B$ exactly one as-yet-untouched string of the form:
$\langle 0^{\widetilde{n}}, \widetilde{w}, b \rangle$,
with $|\widetilde{w}| = \widetilde{n}$ and 
$b \in \{0,1\}$.  Such strings can be found (as, by our choice of the
$n$ of Stage~$i-1$, the simulation we ran during Stage $i-1$ 
cannot touch enough strings to prevent this).

Let $x=0^{n}$.

\noindent
{\it Case 1: $M^{B_{i-1}}(x)$ has a rejecting path.} \quad
We choose one
such path $\alpha$ and freeze it, i.e., we make sure that all queries answered
negatively on $\alpha$ must remain outside of all the $B_{j}$. 
There exist $2^{n}$
triples of the form $\langle 0^{n},y,1 \rangle$ where $|y| = n$, but
only $p(n)$ can occur on $\alpha$. Thus, for some $z$ for which 
$ \langle 0^{n},z,1 \rangle $  is not
queried on $\alpha$, we set
$B_{i}=B_{i-1}\cup \{\langle 0^{n},z,1 \rangle \}$.
Note that Equation~\ref{2} is maintained at this length.

\noindent
{\it Case 2: $M^{B_{i-1}}(x)$ has an accepting path.} \quad
In this case we freeze an accepting path and put some 
$\langle 0^{n},w,0 \rangle $  in $B$.
Note that Equation~\ref{2} is maintained at this length.

\noindent {\it Case 3:  $M^{B_{i-1}}(x)$ has only no-comment paths.}
\quad Note that $M$ is in this case clearly not 
robustly overproductive.  
Add to $B$ exactly one string 
of the form:  $\langle 0^{{n}}, {w}, b \rangle$,
with $|{w}| = {n}$ and 
$b \in \{0,1\}$. 
Note that Equation~\ref{2} is maintained at this length.
\nopagebreak

\nopagebreak\noindent{}{\it End of stage construction}

Assume $A\ored B$ via machine $M$, and 
let $i$ be such that 
this machine was
considered in Stage $i$.  Let $n$ be the input length chosen in
Stage~$i$ and let $x = 0^n$.
Let us consider the implications based 
on which of the three cases applied to 
$M^{B_{i-1}}(x)$.
If Case~1 held, a rejecting path $\alpha$ was frozen, 
and thus, since later stages will not interfere, $\alpha$ 
in fact will be a rejecting path of $M^B(x)$.
By adding 
$\langle 0^{n},z,1 \rangle$  to $B$, we obtained $x \in A$
by Equation~\ref{1}.  There are two cases:
\begin{enumerate}
\item $M^{B}(x)$ has no accepting path.
\item $M^{B}(x)$ has at least one accepting path.
\end{enumerate}
In the former case, $L(M^B) \neq A$,
since $x \in A$, but
$M^B$ does not accept it.
In the latter case, $M^{B}$ on input $x$ fails to be strong, but 
this is 
impossible because $M$ reduces $A$ strongly to $B$ (by the
assumption that $A \ored B$ via $M$).
Case~2 is analogous.
If Case~3 happened, then
$M$ is not robustly overproductive, so 
$A \ored B$ is certainly not implemented by $M$.

So, in all three cases, $M$ does not 
reduce $A$ to $B$ in the sense of $\ored$.  So, due to our 
stage construction, we have that
$A \notored B$.  This completes the proof of 
Part~\ref{p:overpart}.

The proof of 
Part~\ref{p:underpart} of the theorem is very similar to 
that of 
Part~\ref{p:overpart}, so we simply briefly sketch the differences.
At the start of Stage~$i$, where one sets the~$n$ that will be used
in Stage~$i$, we choose~$n$ so large that it not only satisfies 
the conditions used in Part~\ref{p:overpart}, but also so large
that the ``Party Lemma'' applies (see below).  
The three cases of Part~\ref{p:overpart} are then replaced by the 
following argument flow.  If there is 
any length~$n$ string $y$ for which
either 
\begin{enumerate}
\item $M^{(B_{i-1} \cup \langle 0^n,y,1\rangle)}(x)$ does not both
have at least one accepting path and no rejecting paths, or 
\item $M^{(B_{i-1} \cup \langle 0^n,y,0\rangle)}(x)$ does not both
have at least one rejecting path and no accepting paths,
\end{enumerate}
we are easily done, via adding the obvious string to 
taint the reduction.  On the other hand, if 
for each length~$n$ string $y$ we have that
\begin{enumerate}
\item $M^{(B_{i-1} \cup \langle 0^n,y,1\rangle)}(x)$  
has at least one accepting path and no rejecting paths, and
\item $M^{(B_{i-1} \cup \langle 0^n,y,0\rangle)}(x)$ has
at least one rejecting path and no accepting paths,
\end{enumerate}
then (again, assuming $n$ was chosen appropriately large)
by the ``Party Lemma'' 
(\cite[page~104]{cai-gun-har-hem-sew-wag-wec:j:bh2})
there exist strings $y$ and $y'$ (possibly the same string)
of length $n$ such that 
$M^{B_{i-1} \cup \langle 0^n,y,0\rangle
\cup \langle 0^n,y',1 \rangle}(x)$ 
has both accepting and rejecting paths, and thus the machine 
considered at Stage~$i$ in fact is not robustly 
underproductive.  (Of course, we do not in this case add both 
$\langle 0^n,y,0\rangle$ and 
$\langle 0^n,y',1 \rangle$ to our oracle as this 
would taint the promise of Equation~\ref{2}.  Rather, the 
very fact that {\em some oracle\/} causes 
overproductivity is enough.)  So, in this case, we add any single 
coding string of the form
$\langle 0^n, y'' , b \rangle$, with $b\in \{0,1\}$ and 
$|y''| = n$, to maintain Equation~\ref{2}, and we move on to
the next stage.~\qed


Next we consider the relationship between $\ored$ and $\rsred\!$. Let $M$ be an
NPTM\@. By interchanging the accept and the reject states of $M$ we get a new
NPTM machine 
$N$ such that $\lrej(M)=L(N)$. If $M$ is robustly strong, we have 
 $L(M^{A})=\overline{L(N^{A})}$
for every oracle $A$. The pair $(M,N)$ 
is what
Hartmanis
and Hemachandra~\cite{har-hem:j:rob} call 
a robustly complementary pair of machines.
For such a pair, the following is known.
\begin{theorem}{\cite{har-hem:j:rob}} \label{hh90}
\quad
If (M,N) is a robustly complementary pair of machines, then
$$(\forall A)[L(M^{A}) \in {\rm P}^{{\rm SAT} \oplus A}].$$
\end{theorem}

Gavald\`{a} and Balc\'{a}zar~\cite{bal-gav:j:rob}
noted that,
in view of the preceding discussion,
one gets as an immediate corollary the 
following.
\begin{corollary}{\cite{bal-gav:j:rob}}  \label{c:foonew}
\quad
$(\forall A,B)[A \rsred B \longrightarrow A\in {\rm P}^{{\rm SAT} \oplus B}]$.
\end{corollary}

In fact,
the proof of Theorem~\ref{hh90} still works if $M$ is an underproductive
machine reducing $A$ to $B$. Thus, we have the following.

\begin{theorem} \label{uh}
$(\forall A,B)[A \ured B \longrightarrow A\in {\rm P}^{{\rm SAT} \oplus B}]$.
\end{theorem}

Not only is the proof of Theorem~\ref{hh90} not valid for $\ored$, but 
indeed the 
statement of Theorem \ref {uh} with $\ured$ replaced by $\ored$ is outright
false. This follows as a corollary to a 
proof of Crescenzi and 
Silvestri~\cite[Theorem~3.1]{cre-sil:c:sperner}
in which they give a very nice application of Sperner's Lemma.

\begin{theorem}
\label{oh}
$(\exists A,E)[A \ored E \wedge A \not\in {\rm P}^{{\rm SAT} \oplus E}]$.
\end{theorem}
\sproof \quad  
The proof follows from a close inspection of the proof of 
\cite[Theorem~3.1]{cre-sil:c:sperner}.
Crescenzi and Silvestri prove the existence of a 
$\sigmastar$-spanning pair 
$(N_{0},N_{1})$ of machines and the existence of an oracle $E$ with the 
property that
$(\forall \mbox{ 0-1 valued function } 
f\in \fp^{{\rm SAT} \oplus E})(\exists x)
[x \not\in L(N_{f(x)}^{E})]$.
We need some preliminaries. The {\it standard triangulation of size $n$} is the
triangle, $\Delta$, 
in a Euclidean $x,y$-plane with the corners $A=(0,0),B=(n-1,0)$ 
and $C=(0,n-1)$ that is triangulated by the lines $x=i,(i=0,\ldots, n-2),
y=i,(i=0,\ldots, n-2)$ and $x+y=i,(i=1,\ldots,n-1)$. The vertices of this triangulation are exactly the points $(i,j)$ with natural $i$
and $j$, and $i+j\leq n-1$. 
A {\it coloring} of the standard triangulation is a mapping that 
associates with each vertex one of three given colors,
1,~2, and~3. Such a coloring is
called {\it c-admissible}
if the vertices on the border of $\Delta $ satisfy a certain
condition pair, namely the following.
\begin{enumerate}
\item $A$, $B$, and $C$ are colored 1, 2, and 3 respectively.
\item If $P$ is a vertex on a side joining the corners colored $i$ and $j$,
then $P$ must be colored by $i$ or $j$.
\end{enumerate}
Sperner's Lemma guarantees
that each c-admissible coloring of a standard triangulation contains at least 
one three-colored
triangle.

A standard triangulation of size $n$ has $\frac{n(n+1)}{2}$ vertices. 
So a coloring can be encoded by $n(n+1)$ bits,
since a color can easily be encoded using two bits. If $l_{n}$ is 
the smallest
natural number $l$ such that $n(n+1)\leq 2^{l}$, and if a certain ordering
of the vertices of $\Delta$ is fixed, for instance $(0,0),(1,0),\ldots,
(n-1,0),(0,1),(1,1),\ldots,(n-2,1),\ldots,(0,n-1)$, then a coloring 
certainly can always be
encoded by the first $n(n+1)$ bits of a word of length $2^{l_{n}}$, 
which means by a subset $U$ of $\{0,1\}^{l_{n}}$. In order to 
recover the
coloring from $U$, assume $\{0,1\}^{l_{n}}$ to be ordered in the usual 
lexicographic way---$0^{l_n} < 0^{l_n - 1} 1 <\ldots <1^{l_n}$---and 
determine the values of the
characteristic function of $U$ for the first $n(n+1)$ words.
So, the first two values determine the color of $(0,0)$, the next two values 
determine the color of $(1,0)$, and so on.


Let $s_n=\mbox{max}\{k\condition k(k+1)\leq 2^n\}.$
Now the nondeterministic oracle machines $N_{0}$    and $N_{1}$ are defined 
in such 
a way that for arbitrary oracle $C$ and input $x$:
\begin{enumerate}
\item $N_{0}^{C}(x)$ has accepting paths if and only if 
$C\cap \Sigma^{|x|}$ encodes a coloring 
of a standard triangulation of size $s_{|x|}$
containing a
3-colored triangle.
\item $N_{1}^{C}(x)$ has accepting paths if and only if
$C\cap \Sigma^{|x|}$ encodes a non-c-admissible 
coloring of a standard triangulation of size $s_{|x|}$.
\end{enumerate}

Clearly, both machines work in polynomial time. We define a new 
machine $N$
that, on input $x$, nondeterministically transfers the input to both $N_{0}$ 
and 
$N_{1}$
and
\begin{enumerate}
\item On a given path simulating a 
path of $N_1$,
$N$ ends in the state {\bf acc} if $N_{1}$ accepts and
otherwise ends in the no-comment state.
\item On a given path simulating a path
of $N_0$, $N$ ends in the state {\bf rej} if 
$N_{0}$ accepts and otherwise ends in the no-comment state.
\end{enumerate}
By Sperner's Lemma, $(N_{0},N_{1})$ is a robustly 
$\sigmastar$-spanning pair,
and this means that $N$ is robustly overproductive.

The oracle $E$ is defined by diagonalization such that for the $i$th 
deterministic 
polynomial-time oracle machine $T_{i}$ and a suitably 
chosen $n_{i}$ it holds that:
\begin{enumerate}
\item $T_{i}^{{\rm SAT}\oplus E}(0^{n_{i}})=1$ and  
$E\cap \Sigma^{n_i}$ 
encodes
a c-admissible coloring of a standard triangulation of 
size  $s_{n_{i}}$, or
\item $T_{i}^{{\rm SAT}\oplus E}(0^{n_{i}})=0$ and 
$E\cap \Sigma^{n_i}$
encodes
a coloring of a standard triangulation of size $s_{n_{i}}$ without 
three-colored triangles.
\end{enumerate}

The crucial point is that $E$ is in fact constructed in such a way that it 
does not
encode colorings that both contain a three-colored triangle and are 
non-c-admissible. 
The fact that such an $E$ can be constructed
follows from~\cite[Remark~1]{cre-sil:c:sperner}. 
This has the important 
consequence
that $N$ with oracle $E$ is strong. Thus, defining $A=L(N^{E})$, we have
$A\ored E$ via $N$, since $N$ is robustly overproductive.
Finally, we observe that $A\not\in {\rm P}^{{\rm SAT}\oplus E}$. 
Why?  If not, then
we have
some $T_{i}$ with $A=L(T_{i}^{{\rm SAT}\oplus E} )=L(N^{E})$.  However, 
from the 
definition
of $N$ and $E$ (in particular in the stage where $T_{i}$ was considered) 
we 
conclude $0^{n_{i}}\in L(T_{i}^{{\rm SAT}\oplus E} )\longleftrightarrow 
0^{n_{i}}\not\in L(N^{E})$. This contradiction shows 
$A\not\in {\rm P}^{{\rm SAT}\oplus E}$.~\qed

As mentioned
earlier, Theorem~\ref{oh} follows from the {\it proof\/} of
\cite[Theorem~3.1]{cre-sil:c:sperner}, but not from the theorem itself.

The preceding two theorems have the consequence of showing a deep
asymmetry between $\ored$ and $\ured\!$.  This
asymmetry---that 
$\boldored \neq \boldturingred$, yet to prove the analog 
for $\boldured$ would resolve the $\pisnotnp$ question---contrasts with 
the seemingly
symmetric definitions of these two notions.  We  now turn to 
some results that will lead to the proof of this asymmetry.
\begin {theorem}
\label{oud}
\quad Overproductive and underproductive reductions differ
in such a way that
$$\boldored \not\subseteq \boldured\!.$$
\end{theorem}

\sproof
By Theorem~\ref{oh} we have sets $X$ and $Y$ such that 
$$X \ored Y \wedge X \not\in {\rm P}^{{\rm SAT} \oplus Y},$$
and because of Theorem \ref{uh} for these sets the statement
$$X\ured Y \dann X \in {\rm P}^{{\rm SAT}\oplus Y}$$ 
is true. The conjunction of 
these
two statements is equivalent to 
$$X\ored Y \wedge X \notured Y \wedge X \not\in {\rm P}^{{\rm SAT}\oplus Y},$$
from which we conclude $\boldored \not\subseteq \boldured$.~\qed 

An immediate consequence of Theorem~\ref{oud} is the following.

\begin{theorem} \label{rsod}
$\boldrsred \neq \boldored\!$.
\end{theorem}

We conjecture that Theorem~\ref{rsod} can be stated in the much stronger
form of Theorem~\ref{rec}, where $\snred$ is replaced with $\ored\!$.
From Theorem~\ref{rsod}, it follows that $\boldored \neq \boldturingred\!$. 
It is 
interesting to note that, although we know that $\boldored$ and 
$\boldturingred$
 differ,
it may be
extremely hard to prove them to differ with a sparse set on the 
right-hand side. More precisely, we have the following.

\begin{theorem} \label {sprhs}
$(\exists B\in {\rm SPARSE})
[\red{\rm T}{\rm O}{B} \neq \red{\rm T}{\rm p}{B}] \dann
{\rm P} \neq \np$.
\end{theorem}

\sproof \quad 
Assume $\pisnp$ and $A\ored B$ with a sparse set $B$. 
This certainly implicitly gives a robustly $\sigmastar$-spanning
pair $(N_{0},N_{1})$ of machines and, say, $A=L(N_{1}^{B})$.
By \cite[Theorem~2.7]{har-hem:j:rob}, there exists a 
0-1 function $b\in 
\fp^{{\rm SAT}\oplus B}$ 
such that $(\forall x)[x\in L(N_{b(x)}^{B})]$. 
Since $N$ is strong for $B$, the sets
$L(N_{0}^{B})$ and $L(N_{1}^{B})$ are disjoint. Hence $x \in L(N_{0}^{B})$
if and only if $x \not\in L(N_{1}^{B})$, from which we conclude
$$x\in A \longleftrightarrow b(x)=1.$$
From the assumption $\pisnp$,  it follows that $b\in \fp^{B}$ and thus
$A\turingred B$.~\qed

The fact  $\boldored \neq \boldturingred\!$, stated above, 
sharply contrasts with the following.

\begin{theorem}
\label{upd}
$\boldured \neq \boldturingred \dann \pisnotnp$. 
\end{theorem}
\sproof
Assume $\pisnp$ and $A \ured B$. By Theorem \ref{uh} we have $ A \in 
{\rm P}^{{\rm SAT}\oplus B}$, 
and because of our $\pisnp$ 
assumption, this means $A \turingred B$.~\qed

Theorem~\ref{upd} strengthens in two ways the statement, 
noted by 
Gavald\`{a} and Balc\'{a}zar~\cite{bal-gav:j:rob},
that if $\rsred$ differs from $\turingred$
{anywhere on the recursive sets} then $\pisnotnp$.
In particular, we have these two improvements of 
that statement of 
Gavald\`{a} and Balc\'{a}zar:
(a)~we improve from $\rsred$ to $\ured$, and
(b)~we remove the ``on the recursive sets'' scope restriction.

Below, we use $X\not\subset Y$ to denote that it is not the case that
$X\subsetproper Y$.

\begin{corollary}~
\begin{enumerate}
\item 
$\boldured \not\subset
\boldored \dann \pisnotnp$.
\item $\boldured \not= \boldrsred \dann \pisnotnp$.
\end{enumerate}
\end{corollary}
\sproof
By Theorem \ref {upd}, $\pisnp$ implies $\boldured =\boldturingred$ and thus
$\boldured \subseteq \boldored\!$. In light 
of Theorem~\ref{oud}, we even have
$\boldured \subsetproper \boldored$.~\qed

So proving $\boldured \neq \boldturingred, \boldured \neq \boldrsred$, or
$\boldured \not\subset
\boldored$ amounts to proving 
$\pisnotnp$. In particular, we cannot hope to strengthen Theorem~\ref{rec}
so that it is valid for $\ured$ rather than $\snred$.

Although we
know that  $\boldored \not\subseteq \boldured\!$, 
it is also difficult to show
that they differ with respect to
a sparse set on the right hand side, because we have
$$(\exists B \in {\rm SPARSE})[\red{\rm T}{\rm O}{B}\not\subseteq 
\red{\rm T}{\rm U }{B} ] \dann
\pisnotnp,$$
which is a consequence of Theorem~\ref{sprhs}.
\begin{theorem}~
\begin{enumerate}
\item $\boldrsred \neq \boldturingred \dann \pisnotnp.$
\item $\boldrsred = \boldturingred \dann {\rm P}=\np \cap \conp.$
\end{enumerate}
\end{theorem}
\sproof
Let $ {\rm P}=\np$. Then from $A\rsred B$ we have, by 
Corollary~\ref{c:foonew},
$ A \in {\rm P}^{{\rm SAT} \oplus B}={\rm P}^{B}$, i.e., $A\turingred B$.
If $\rsred =\turingred$, then their zero degrees coincide, so
$ {\rm P}=\np \cap \conp.$~\qed

\section{Overproductive Reductions
and the Classic Hardness Theorems}\label{s:hardness-theorems}





The polynomial hierarchy is defined as follows:
(a)~$\sigmazero = \p$;
(b)~for each $i\geq 0$, 
$\sigmaiplusone = \np^{\sigmai}$;
(c)~for each $i\geq 0$, 
$\pii = \{L \condition \overline{L}\in \sigmai\}$; and
(d)~$\ph = \cup_{i \geq 0} \sigmai$~\cite{sto:j:poly}.
$\thetatwo = \{ L \condition L \leq_{\rm tt}^{\rm p} 
\sat\}$ (see~\cite{wag:j:bounded}),
where $\leq_{\rm tt}^{\rm p}$ denotes polynomial-time
truth-table reduction.
ZPP denotes expected 
polynomial time~\cite{gil:j:probabilistic-tms}.
It is well-known that
$\np\subseteq \thetatwo \subseteq \p^{\np} \subseteq 
\zpp^{\np} \subseteq \sigmatwo$.

It is very natural to ask
whether the existence 
of sparse hard or complete sets with respect to our new
reductions would imply collapses of the 
polynomial hierarchy similar to those  
that are known to hold for $\turingred\!$.
That is, are our reductions useful in extending the 
key standard results?  To study this question, we 
must first briefly review what is known regarding 
the consequences of the existence of sparse 
NP-hard sets.  The classic result in this direction
was obtained by Karp and Lipton, and more recent
research has yielded three increasingly strong
extensions of their result. 

\begin{theorem} \label{t:fourparts}
\begin{enumerate}

\item \label{p:karp-lipton}
\cite{kar-lip:c:nonuniform} \quad
$\np \subseteq \red{\rm T} {\rm p} {\rm SPARSE} \dann \ph \subseteq 
\sigmatwo.$

\item \label{p:kamper}
(implicit in \cite{kar-lip:c:nonuniform}, 
see~\cite{lon-sel:j:sparse} and the discussion
in~\cite{hem-hoe-nai-ogi-sel-thi-wan:j:np-selective};
explicit in~\cite{aba-fei-kil:j:hide,kae:j:nup}) \quad  
$\np \subseteq \red{\rm T} {\rm RS} {\rm SPARSE} \dann \ph \subseteq 
\sigmatwo.$

\item \label{Now-just-part:koewat}
\cite{koe-wat:cOutBySicompElecPub:new-collapse} \quad
$\np \subseteq \red{\rm T} {\rm RS} {\rm SPARSE} \dann \ph \subseteq \zppnp.$

\item \label{oncemore:koewat}
\cite{koe-wat:cOutBySicompElecPub:new-collapse} \quad
If $A$ is self-reducible and 
$A\in (\np^{B} \cap \conp^{B})/\poly$,
then $\zppnp^{A} \tm \zppnp^{B}$.

\item \label{p:ks}
\cite{koe-sch:b:high-for-np} \quad
If $A$ has self-computable witnesses and 
$A\in (\np^{B} \cap \conp^{B})/\poly$,
then $\zppnp^{A} \tm \zppnp^{B}$.

\end{enumerate}
\end{theorem}

We mention that 
K\"obler and Watanabe~\cite{koe-wat:cOutBySicompElecPub:new-collapse} 
state part~\ref{Now-just-part:koewat}
in the form
$\np \tm (\np \cap \conp)/\poly \dann \ph \tm \zppnp$,
which is equivalent to the statement of 
part~\ref{Now-just-part:koewat}  
in light of Theorem~\ref{spcl}. Both part~\ref{oncemore:koewat} and 
part~\ref{p:ks} extend part~\ref{Now-just-part:koewat}. 


It remains open whether 
parts~\ref{Now-just-part:koewat}, \ref{oncemore:koewat},
or~\ref{p:ks} of 
Theorem~\ref{t:fourparts} can be extended from 
robustly strong reductions to overproductive reductions.
However, as Theorem~\ref{theorem-kl-NEW} we extend part~\ref{p:kamper}
of Theorem~\ref{t:fourparts}
to overproductive reductions.  
As a consequence, there is at the present time no 
single strongest theorem on this topic;
Theorem~\ref{theorem-kl-NEW} 
seems to be incomparable 
in strength relative to either of the final two parts
of Theorem~\ref{t:fourparts}.


\begin{theorem}\label{theorem-kl-NEW}
$\np \subseteq \red{\rm T}{\rm O}{\rm SPARSE} \dann \ph \subseteq 
\sigmatwo$.
\end{theorem}
\sproof \quad
Our proof will in effect extend the approach 
of Hopcroft's~\cite{hop:c:recent} proof of the 
Karp-Lipton Theorem (Theorem~\ref{t:fourparts},
part~\ref{p:karp-lipton}) in a way that allows the proof to 
work even when the machine involved is one implementing
an overproductive reduction.  We will centrally use
the fact that such machines are also underproductive for the 
specific set to which the reduction maps.

Assume 
$\np \subseteq \red{\rm T}{\rm O}{\rm SPARSE}$.
Then there is a sparse set $S$ such that 
$\sat \leq_{\rm T}^{\rm O} S$, and let $M$
be a machine certifying the reduction.  
That is, NPTM $M$ is robustly overproductive, 
$M^S$ is underproductive, and $\sat = L(M^S)$.
Let $p_S$ bound the 
sparseness of $S$, i.e., 
for each $m$, $|| S^{\leq m}|| \leq p_S(m)$.

Let $L$ be an arbitrary $\pitwo$ set.
So for some NPTM $N$ we have  $\overline{L} = 
L(N^{\sat})$.  We will describe a $\sigmatwo$
algorithm for $L$.  Say the runtime of $N$ (respectively, $M$)
is upper-bounded (without loss of generality, for all 
oracles)
by $p_N$ (respectively, $p_M$).  
Our $\sigmatwo$ algorithm will be implemented 
by an NPTM $\widehat{N}$ with SAT as its oracle.  
Since SAT is NP-complete, we will act as if 
$\widehat{N}$ had two different NP sets 
($A$ and $B$, defined below) as 
its oracle;  implicitly, each when called is implemented 
via a reduction to SAT\@.

We now describe $\widehat{N}$.  For each $y$ that is a 
boolean formula with at least one variable,
let $y_T$ denote $y$ with its first variable set to true,
and let 
$y_F$ denote $y$ with its first variable set to false.
Let the function ``$\coding$'' be such that given any finite set 
$R$, $\coding(R)$ is a standard, easily decodable
encoding of $R$.  For a computation path $\rho$ 
(of some NPTM 
implementing an overproductive reduction), 
let $\outcome(\rho)$ denote the outcome of the
path~$\rho$ (which will be one of {\bf acc}, {\bf rej}, 
or~{\bf{}?}).~~On 
input $x$, $|x|= n$, 
$\widehat{N}$ nondeterministically guesses each subset, $R$,
of $\Sigma^{\leq p_M(p_N(n))}$ containing at most 
$p_S(p_M(p_N(n)))$ elements.  
$\widehat{N}$
then asks
$\pairs{x,\coding(R)}$ to the NP set $A$ implicitly 
defined by the following.
$\pairs{x, H} \in \overline{A}$ if and only if 
there is an $R$, with $H = \coding(R)$ such 
that, for each string $y$ satisfying $|y| \leq p_N(|x|)$,
the following conditions hold:
\begin{enumerate}
\item if $y$ is a legal formula with at least one
variable then\\
$(\forall \rho_1 : \rho_1$ is a path of $M^R(y)$ and 
$\outcome(\rho_1) \in 
\{ \mbox{\bf{}acc},\mbox{\bf{}rej} \} )$\\
$(\forall \rho_2 : \rho_2$ is a path of $M^R(y_T)$ and 
$\outcome(\rho_2) \in 
\{ \mbox{\bf{}acc},\mbox{\bf{}rej} \} )$\\
$(\forall \rho_3 : \rho_3$ is a path of $M^R(y_F)$ and 
$\outcome(\rho_3) \in 
\{ \mbox{\bf{}acc},\mbox{\bf{}rej} \} ) $\\
$[
\outcome(\rho_1) = \mbox{\bf{}acc} \iff
\outcome(\rho_2) = \mbox{\bf{}acc} \lor
\outcome(\rho_3) = \mbox{\bf{}acc}
]
$, and

\item
if $y$ is a legal formula with no 
variables then \\
$(\forall \rho : \rho$ is a path of $M^R(y))$\\
$[ 
(y \equiv  \true \oldimplies \outcome(\rho) \neq  \mbox{\bf{}rej})
\land 
(y \equiv \false \oldimplies \outcome(\rho) \neq  \mbox{\bf{}acc})
]$.
\end{enumerate}
Crucially, note that for each $x$ it will
hold that for at least one $R$ the query
$\pairs{x,\coding(R)}$ that $\widehat{N}$ asks will 
be such that 
$\pairs{x,\coding(R)} \not\in A$.  
For each $R$ satisfying 
$\pairs{x,\coding(R)} \not\in A$, note that by the 
definition of $A$ we have that
(a)~$M^R(y)$ is 
underproductive for all $y$ satisfying
$|y| \leq p_N(|x|)$, and 
(b) $\sat^{\leq p_N(n)} = {\left( L(M^R) \right)}^{\leq p_N(n)}$.
(Additionally, recall that $M$ is robustly overproductive.)
The key point here is that if $M^R$ actually overproduces
(has both accepting and rejecting paths) on some
$y$ with 
$|y| \leq p_N(|x|)$, the test will be effected by this 
in such a way that we will have 
$\pairs{x,\coding(R)} \in A$.  In particular, the set
$\overline{A}$ is essentially, regarding internal nodes of $\sat$'s 
disjunctive self-reducibility tree, testing that 
each triple of non-{\bf{}?}~outputs, one each from
a node and its children, is such that the three outputs
are consistent with a correct self-reduction; if a machine 
overproduces anywhere (in the range considered) 
it will fail this test.

$\widehat{N}$, on each guessed path (that is, each 
guess of $R$) that gets the answer
$\pairs{x,\coding(R)} \in A$ simply rejects.
$\widehat{N}$, on each guessed path that gets the answer
$\pairs{x,\coding(R)} \not\in A$ asks the query
$\pairs{x,\coding(R)}$ to the set $B\in \np$,
and accepts if and only if the answer to this 
query is 
$\pairs{x,\coding(R)}\not\in B$. $B$ is defined as 
follows.   
$\pairs{x, H} \in {B}$ if and only if 
there is an $R$ with $H = \coding(R)$ such 
that nondeterministically simulating 
$N^{L(M^R)}(x)$ yields at least one accepting 
path (of $N$),
where by ``simulating'' we mean simulating $N$ and,
each time an oracle call $w$ is made to 
$L(M^R)$, nondeterministically guessing a path 
$\rho'$ of $M^R(w)$ and 
(a)~continuing the simulation of $N$ with 
the answer yes (respectively, no) if 
the outcome of $\rho'$ is 
{\bf{}acc} (respectively, {\bf{}rej}),
and (b)~halting and rejecting 
(on the current path---recall that $N$ is a 
standard Turing machine whose paths thus each 
either accept or reject, and by definition
the machine accepts an input exactly if there
is some accepting path on that input)
if $\rho'$
has~{\bf{}?}~as its outcome.
The crucial point here is that, for those $R$ on which
it is actually called on actual runs 
of $\widehat{N}$, $B$'s use of $R$ will correctly simulate
$\sat$.

We have shown that each $\pitwo$ set has a $\sigmatwo$ 
algorithm, and thus have proved our theorem.~\qed


The above proof does not work for the case of underproductive
reductions, and indeed it remains open whether
Theorem~\ref{theorem-kl-NEW} can in some way 
be extended to underproductive
reductions.  An analog for strong nondeterministic 
reductions is implicitly known, but has a far weaker
conclusion.

\begin{theorem} 
(implicit in~\cite{koe-wat:cOutBySicompElecPub:new-collapse})\quad 
$\np \subseteq \red {\rm T}{\rm SN}{\rm SPARSE} \dann \ph \tm 
\zpp^{\sigmatwo}.$
\end {theorem}
\sproof
We start out from  Yap's theorem~\cite{yap:j:advice}
in its strengthened form found by 
K\"obler and Watanabe~\cite{koe-wat:cOutBySicompElecPub:new-collapse}, 
namely,
$\conp \subseteq \np/\poly \dann \ph \subseteq \zpp^{\sigmatwo}$.
The following two statements show that  the theorem to be proved is
simply an equivalent reformulation of Yap's theorem:
$\np \subseteq \red {\rm T}{\rm SN}{\rm SPARSE}\gdw
\conp \subseteq \red {\rm T}{\rm SN}{\rm SPARSE}$ and
(recalling Theorem~\ref{spcl})
$\conp \subseteq
\red {\rm T}{\rm SN}{\rm SPARSE}\dann\conp \subseteq \np/\poly$.~\qed

In contrast with the above results regarding 
sparse hard sets for NP, in the case of sparse 
{\em complete\/} sets for NP we have just as strong 
a collapse for $\snred\!$-reductions as we 
have for $\turingred\!$-reductions.

\begin{theorem}\label{t:mkk}
\cite{kad:j:pnplog} \quad 
$\np \subseteq \red{\rm T}{\rm SN}{\rm SPARSE \cap \np}
\dann \ph =
\thetatwo$.
\end{theorem}

\sproof \quad
Kadin~\cite{kad:j:pnplog} proved:
If there exists a set $S\in \np \cap \sparse$ such that $\conp \subseteq
\np^{S}$, then $\ph = \thetatwo$.
This is equivalent to 
Theorem~\ref{t:mkk}, because $\conp \subseteq \np^{S}$ is 
equivalent to $\np \subseteq \np^{S} \cap \conp^{S}$, and this, in turn,
is equivalent to
$\np \subseteq \red{\rm T}{\rm SN}{S}$.~\qed

As mentioned earlier,
we leave as an open problem whether one 
can establish the collapse $\ph \subseteq \sigmatwo$
(or, better still, $\ph \subseteq \zppnp$)
under the assumption $\np \subseteq 
\red{\rm T}{\rm SN}{\rm SPARSE}$, 
or 
even under the stronger assumption that $\np \subseteq 
\red{\rm T}{\rm U}{\rm SPARSE}$.
We conjecture that no such 
extension is possible.

\section{Conclusions and Open Problems}\label{s:local-vs-global}

We mention as open problems the issues of 
finding equivalent conditions for $\boldured \subseteq \boldored\!$,
$\boldrsred = \boldturingred\!$, and $\boldured =\boldrsred\!$.

Define the runtime of a nondeterministic machine on a given
input to be the length of its longest computation path.
(Though in most settings this is just one of a few equivalent 
definitions, we state it explicitly here as for the about-to-be-defined
notion of local-polynomial machines, it is not at all clear
that this equivalence remains valid.)
Recall that we required that NPTMs be such that 
for each NPTM, $N$, it holds that there exists a polynomial
$p$ such that, for each oracle $D$, the runtime of $N^D$ is 
bounded by $p$.  Call such a machine ``global-polynomial''
as there is a polynomial that globally bounds its
runtime.
Does this differ from a requirement 
that for a machine $N$ it holds that,
for each oracle $D$, there is a polynomial
$p$ (which may depend on $D$) such that the runtime of $N^D$ 
is bounded by $p$?  Call such a machine ``local-polynomial''
as, though for every oracle it runs in polynomial time, the 
polynomial may depend on the oracle.

In general, these notions do differ, notwithstanding the common wisdom
in complexity theory that one may ``without loss of generality''
assume enumerations of machines come with attached clocks independent
of the oracle.  (The subtle issue here is that the notions in fact
usually do
not differ on enumerations of machines that 
will be used with only one oracle.)
The fact that they in general differ is made clear by the
following theorems.  These theorems show that there is a language
transformation that can be computed by a local-polynomial machine,
yet each global-polynomial machine will, for some target set, fail almost 
everywhere to compute the set's image
under the language transformation.  We write $A =^* B$ if $A$ and 
$B$ are equal almost everywhere, i.e., if $(A-B) \cup (B-A)$ is 
a finite set.

\begin{theorem}
There is a function $f_N: 2^{\sigmastar} \rightarrow 2^{\sigmastar}$
(respectively, $f_D: 2^{\sigmastar} \rightarrow 2^{\sigmastar}$)
such that
\begin{enumerate}
\item there is a nondeterministic 
(respectively, deterministic) local-polynomial 
Turing machine $\widehat{M}$ such that for each oracle $A$ it holds
that $L(\widehat{M}^A) = f_N(A)$
(respectively, $L(\widehat{M}^A) = f_D(A)$), 
and
\item for each NPTM, i.e., each nondeterministic global-polynomial
Turing machine $M$ 
(respectively, DPTM, i.e., each deterministic global-polynomial
Turing machine $M$)
it holds that there is a set
$A\subseteq \sigmastar$ such that
$L(M^A) =^* \overline{f_N(A)}$
(respectively,
$L(M^A) =^* \overline{f_D(A)}$).
\end{enumerate}
\end{theorem}

Though this claim may at first seem counterintuitive,
its proof is
almost immediate if one is given $f_N$ and $f_D$, and so we 
simply give 
functions $f_N$ and $f_D$ satisfying the theorem.
In particular, we can use 
$f_N(A) = \{x \condition (\exists y) [ ( |y| \leq \log |x|)
\land 
(y$ is the lexicographically first string in $A) \land
(\exists z) [ |z| = |x|^{|y|} \land xz \in A]]\}$
and 
$f_D(A) = \{x \condition (\exists y) [ ( |y| \leq \log |x|)
\land 
(y$ is the lexicographically first string in $A) \land
(\exists z) [ (z$ is one of the $|x|^{|y|}$ lexicographically
smallest length $|x|^{|y|}$ strings in $\sigmastar)
\land xz \in A]]\}$.

The difference between global-polynomial machines and local-polynomial
machines in general mappings, as just proven, may make one wonder
whether the fact
that robust strong reduction is defined in terms of global-polynomial
(as opposed to local-polynomial) machines makes a difference and, if
so, which definition is more natural.  Regarding the former issue, we
leave it as an open question.   (The above theorems do 
not resolve this issue, as they deal with
language-to-language transformations defined 
specifically over all of $2^{\sigmastar}$, but in contrast a robustly strong
reduction must accept a specific language only for one oracle, 
and for all others merely has to be underproductive and
overproductive, plus it must have the global-polynomial property.)
That is, the open question is: Does there exist a pair of sets $A$ and
$B$ such that $A \notrsred B$ (which by definition involves a
global-polynomial machine) and yet
there exists a nondeterministic local-polynomial
Turing machine $N$ such that $L(N^B) = A$ and $(\forall
D \subseteq \sigmastar)[N^D$ is both underproductive and
overproductive$]$?  Regarding the question of naturalness, this is a
matter of taste.  However, we point out that the global-polynomial
definition is exactly that of Gavald\`{a} and
Balc\'{a}zar~\cite{bal-gav:j:rob}, and that part~\ref{p:rs} of
Theorem~\ref{spcl}, Gavald\`{a} and
Balc\'{a}zar's~\cite{bal-gav:j:rob} natural characterization of
robustly strong reductions to sparse sets in terms of the complexity
class $(\np\inter\conp)/{\rm poly}$, seems to depend crucially on the
fact that one's machines are global-polynomial.  

On the other hand Theorem~\ref{theorem-kl-NEW}, though its proof seems
on its surface to be dependent on the fact that $\leq_{\rm
  T}^{\rm O}$ is defined via global-polynomial machines, in fact
remains true even if $\leq_{\rm T}^{\rm O}$ is redefined via
local-polynomial machines.  The trick here is that we 
modify the proof to clock the key
local-polynomial machine (implementing the 
overproductive reduction) with the clock that applies for the sparse
oracle to which the reduction actually reduces it, and then in the
simulations of the proof if we detect that a path is about to exceed
that clock, we know that we are dealing with a bad oracle $R$, and so in
our simulation of that too-long path we truncate the path and ``cap''
it with two leaves, one an {\bf acc} leaf and one a {\bf rej} leaf.
This, in effect, rules out that potential oracle, as it will seem to
be overproductive.

{\samepage
\begin{center}
{\bf Acknowledgments}
\end{center}
\nopagebreak
\indent
We are grateful to Yenjo Han, who first noticed the 
error in the proof of Gavald\`{a} and Balc\'{a}zar.
We thank Edith Hemaspaandra,
J\"org Rothe, Osamu Watanabe, and the anonymous referees
for 
many helpful comments and corrections.
We particularly thank
Osamu Watanabe for 
pointing out that a step of the proof of Theorem~\ref{notrec}
uses the self-reducibility of $\et$.

}


\end{document}